\documentclass[fleqn,usenatbib]{mnras}

\usepackage{newtxtext,newtxmath}
\usepackage[T1]{fontenc}
\usepackage{ae,aecompl}
\usepackage[table]{}

\usepackage{graphicx}	% Including figure files
\usepackage{amsmath}	% Advanced maths commands
\usepackage{hyperref}
\usepackage{tablefootnote}
\usepackage{wasysym}
\usepackage{float}
\usepackage{tabularx}
\usepackage{comment}

\title[Discovery of GPX-1\,b]{Discovery of a young low-mass brown dwarf transiting a fast-rotating \\ F-type star by the Galactic Plane eXoplanet (GPX) survey}

\author[Benni et al.]{\parbox{\textwidth}{P.~Benni$^{1}$\thanks{E-mail: pbenni@verizon.net (PB);  burdanov.art@gmail.com (AB)},
A.~Y.~Burdanov$^{2,3}$\color{blue}\footnotemark[1]\color{black},
V.~V.~Krushinsky$^{4}$,
A.~Bonfanti$^{5,6}$,
G.~H{\'e}brard$^{7,8}$,
J.~M.~Almenara$^{9}$,
S.~Dalal$^{7}$,
O.~D.~S.~Demangeon$^{10}$,
M.~Tsantaki$^{11}$,
J.~Pepper$^{12}$,
K.~G.~Stassun$^{13}$,
A.~Vanderburg$^{14,15}$,
A.~Belinski$^{16}$,
F.~Kashaev$^{17}$,
K.~Barkaoui$^{18,19}$,
T.~Kim$^{20}$,
W.~Kang$^{20}$,
K.~Antonyuk$^{21}$,
V.~V.~Dyachenko$^{22}$,
D.~A.~Rastegaev$^{22}$,
A.~Beskakotov$^{22,23}$,
A.~A.~Mitrofanova$^{22}$,
F.~J.~Pozuelos$^{18,6}$,
E.~D.~Kuznetsov$^{24}$,
A.~Popov$^{24}$,
F.~Kiefer$^{7}$,
P.~A.~Wilson$^{25,26}$,
G.~Ricker$^{27}$,
R.~Vanderspek$^{27}$,
D.~W.~Latham$^{28}$,
S.~Seager$^{27,2,29}$,
J.~M.~Jenkins$^{30}$,
E.~Sokov$^{23,31}$,
I.~Sokova$^{23,31}$,
A.~Marchini$^{32}$,
R.~Papini$^{33}$,
F.~Salvaggio$^{33}$,
M.~Banfi$^{33}$,
\"O.~Ba\c{s}t\"urk$^{34}$,
\c{S}.~Torun$^{34}$,
S.~Yal\c{c}{\i}nkaya$^{34}$,
K.~Ivanov$^{35}$,
G.~Valyavin$^{22,21,23}$,
E.~Jehin$^{6}$,
M.~Gillon$^{18}$,
E.~Pak{\v s}tien{\. e}$^{36}$,
V.-P.~Hentunen$^{37}$,
S.~Shadick$^{38}$,
M.~Bretton$^{39}$,
A.~W\"{u}nsche$^{39}$,
J.~Garlitz$^{40}$,
Y.~Jongen$^{41}$,
D.~Molina$^{42}$,
E.~Girardin$^{43}$,
F.~Grau~Horta$^{44}$,
R.~Naves$^{45}$,
Z.~Benkhaldoun$^{19}$,
M.~D.~Joner$^{46}$,
M.~Spencer$^{46}$,
A.~Bieryla$^{28}$,
D.~J.~Stevens$^{47,48,49}$,
E.~L.~N.~Jensen$^{50}$,
K.~A.~Collins$^{28}$,
D.~Charbonneau$^{28}$,
E.~V.~Quintana$^{51}$,
S.~E.~Mullally$^{52}$,
C.~E.~Henze$^{30}$
}
}

\date{Affiliations can be found after the references; Accepted to MNRAS in May 2021.}

\pubyear{2020}

\begin{document}
\label{firstpage}
\pagerange{\pageref{firstpage}--\pageref{lastpage}}
\maketitle

\begin{abstract}

We announce the discovery of GPX-1\,b, a transiting brown dwarf with a mass of $19.7\pm 1.6$\,$M_{\mathrm{Jup}}$ and a radius of $1.47\pm0.10$\,$R_{\mathrm{Jup}}$, the first sub-stellar object discovered by the Galactic Plane eXoplanet (GPX) survey. The brown dwarf transits a moderately bright ($V$ = 12.3\,mag) fast-rotating F-type star with a projected rotational velocity $v\sin{\,i_*}=40\pm10$\,km/s. We use the isochrone placement algorithm to characterize the host star, which has effective temperature $7000\pm200$\,K, mass $1.68\pm0.10$\,$M_{\astrosun}$, radius $1.56\pm0.10$\,$R_{\astrosun}$ and approximate age $0.27_{-0.15}^{+0.09}$\, Gyr. GPX-1\,b has an orbital period of $\sim$1.75\,d, and a transit depth of $0.90\pm0.03$\,\%. We describe the GPX transit detection observations, subsequent photometric and speckle-interferometric follow-up observations, and SOPHIE spectroscopic measurements, which allowed us to establish the presence of a sub-stellar object around the host star. GPX-1 was observed at 30-min integrations by TESS in Sector 18, but the data is affected by blending with a 3.4\,mag brighter star 42\,arcsec away. GPX-1\,b is one of about two dozen transiting brown dwarfs known to date, with a mass close to the theoretical brown dwarf/gas giant planet mass transition boundary. Since GPX-1 is a moderately bright and fast-rotating star, it can be followed-up by the means of Doppler tomography.  
\end{abstract}


%%%%%%%%%%%%%%%%%%%%%%%%%%%%%%%%%%%%%%%%%%%%%%%%%%

%%%%%%%%%%%%%%%%% BODY OF PAPER %%%%%%%%%%%%%%%%%%

\section{Introduction}\label{Introduction}

Brown dwarfs (BDs) are sub-stellar objects with masses in the range of $\sim$13-80\,$M_{\mathrm{Jup}}$. Objects within these mass limits are below the hydrogen-burning minimum mass of 0.07-0.08\,$M_{\astrosun}$ ($\sim$80\,$M_{\mathrm{Jup}}$) and cannot sustain thermonuclear fusion of hydrogen or helium \citep{1963ApJ...137.1121K,1963PThPh..30..460H}, but reactions with deuterium and lithium are possible \citep{1997ApJ...491..856B,2000ApJ...542..464C,2014prpl.conf..619C}. The exact mass boundaries depend on the chemical composition of the sub-stellar object \citep{2002A&A...382..563B,2011ApJ...727...57S}. The formation processes of BDs are still debated, but sub-stellar objects above $\sim$13\,$M_{\mathrm{Jup}}$ are generally considered BDs regardless of formation mechanism or current location
\citep{2007IAUTA..26..183B}. It is still not well understood whether BDs are a product of protostellar cloud fragmentation or if they are formed in protoplanetary discs around young stars. Current statistics of the observed properties of BDs indicates that two distinct populations exist: massive BDs ($\gtrsim\,40\,M_{\mathrm{Jup}}$), which are formed similar to binary stars, and low-mass BDs, which are formed as planets (e.g., \citealt{2014MNRAS.439.2781M,2016A&A...588A.144W}). However, this division of BDs into two populations is being challenged with the new discoveries of intermediate-mass BDs \citep{2019AJ....158...38C,2020arXiv200201943C,2019MNRAS.489.5146J,2020AJ....159..151S}.

The relatively large masses and radii of BDs (compared to most planets) should make them readily detectable in radial velocity (RV) and photometric surveys. However, most known BDs\footnote{\url{http://www.johnstonsarchive.net/astro/browndwarflist.html}} have been found as isolated objects, and far fewer BDs are known to be in multiple systems \citep{2011A&A...525A..95S,2019A&A...631A.125K}. Transiting BDs are even rarer, but these systems provide a unique opportunity to probe the properties of these objects by making possible the measurement of their radii, masses, and obliquities, which can give initial glimpses into their dynamical history \citep{2019AJ....157...31Z}. The lack of detections of BDs with orbital periods shorter than 10\,yr around main-sequence stars has been known as the brown dwarf desert \citep{2000PASP..112..137M,2000A&A...355..581H,2006ApJ...640.1051G}. It is associated with the different formation mechanisms of low- and high-mass BDs and with instability of their orbits, where tidal interaction causes the loss of angular momentum and orbit shrinking until the engulfment of a BD by its host star \citep{2002MNRAS.330L..11A, 2002ApJ...568L.117P}. For BDs orbiting late-type stars with a deep convective layer, magnetic braking should speed up the merging. However, more and more detections "populate" the desert (e.g., \citealt{2016A&A...588A.144W,2019A&A...631A.125K}), while BDs with orbital periods shorter than 100\,d still remain quite rare. A detailed understanding of the formation and evolution of BDs orbiting stars is still difficult due to the lack of statistical data.

Exoplanet RV and transit surveys keep expanding the current population of BDs (see the Exoplanet.eu database; \citealt{2011A&A...532A..79S}). Despite the fact that roughly half of the known transiting BDs were discovered with the use of space telescopes, ground-based exoplanet surveys detected a number of the BDs known to transit relatively bright host stars in the $V\leq13$ range: KELT-1 \citep{2012ApJ...761..123S}, HATS-70 \citep{2019AJ....157...31Z}, WASP-30 \citep{2011ApJ...726L..19A}, and WASP-128 \citep{2018MNRAS.481.5091H}. Such systems allow more in-depth follow-up studies compared to systems with fainter host stars.  

The ground-based transit surveys, which discovered the above-mentioned BDs transiting bright stars, have observed a substantial portion of the sky in an attempt to find new transiting planets. However, most of these surveys do not observe very dense parts of the Galactic plane in order to avoid problems associated with blending of the stars. This is especially true for wide-field surveys like WASP that had poor spatial resolution (the WASP image scale is 13.7\,arcsec\,pixel$^{-1}$). Blending complicates the detection of transit signals and can significantly increase the rate of false-positive exoplanet candidates. The Kepler/K2 space missions \citep{Borucki2010,2014PASP..126..398H} had better spatial resolution (4\,arcsec\,pixel$^{-1}$) than most ground-based exoplanet transit surveys, and brought a significant contribution to the known sample by discovering 40 hot Jupiters and 6 BDs \citep{2011A&A...532A..79S}, but Kepler/K2 was limited by the fact that it observed only some parts of the sky. The Transiting Exoplanet Survey Satellite survey (TESS; \citealt{2014SPIE.9143E..20R}) that has been operating since 2018 will observe almost the whole sky, but it also has quite low spatial resolution (20.25\,arcsec\,pixel$^{-1}$). The future PLAnetary Transits and Oscillation of stars mission (PLATO; \citealt{2014ExA....38..249R}) has a resolution of 15\,arcsec\,pixel$^{-1}$ and it will perform long-duration observations of only two sky fields. OGLE (Optical Gravitational Lens Experiment; \citealt{2003AcA....53..291U}) is a large southern sky variability survey, which observes the Galactic Bulge and Disk with an image scale of 0.26\,arcsec\,pixel$^{-1}$ that allows detection of transiting exoplanets in the dense stellar fields (e.g., \citealt{2003Natur.421..507K,2008A&A...487..749P}). However, only small parts of the Galactic Bulge and Disk are observed with a cadence that allows transit detections \citep{2015AcA....65....1U}.

Therefore, there is an opportunity for a dedicated exoplanet survey that will explore the Galactic plane with sufficient spatial resolution and cadence to find new transiting exoplanets. Motivated by this, we initiated the Galactic Plane eXoplanet (GPX) survey. GPX is a multinational project involving a collaboration of amateur and professional astronomers from Europe, Asia, and North America. GPX evolved from the prototype KPS survey \citep{2016MNRAS.461.3854B}, which resulted in a discovery of the transiting hot Jupiter KPS-1\,b \citep{2018PASP..130g4401B}. The main goal of GPX is to survey high-density star fields of the Galactic plane with moderately high image resolution of 2 arcsec~pixel$^{-1}$ in order to find new transiting gas giants. As a secondary goal, the GPX survey performs search for peculiar variable stars (e.g., see \citealt{2020MNRAS.493.5208K}). In this paper, we present the discovery of a transiting BD with a mass of $19.7\pm 1.6$\,$M_{\mathrm{Jup}}$ and a radius of $1.47\pm0.10$\,$R_{\mathrm{Jup}}$ orbiting F2 star 2MASS 02332859+5601325 (henceforth referred to as the GPX-1) with a period of $\sim$1.75\,d. GPX-1 is located 42\,arcsec from a bright ($V \sim$9\,mag) star HD\,15691 (see~Fig.~\ref{fig_Map_and_LC}). The proximity of this bright star dilutes the transit depth when photometric observations are done with a low spatial resolution. 

The rest of the paper is organized as follows: Section~\ref{sec:obs} describes the GPX discovery wide-field photometry, high-precision photometric follow-up observations, speckle-interferometric observations, spectroscopic follow-up,
and data reduction. Section~\ref{sec:analysis} is devoted to the joint analysis of  GPX-1 and its transiting BD. In Section~\ref{sec:discuss} we discuss obtained results, and we summarise our findings in Section~\ref{sec:conclusions}.

\section{DISCOVERY AND FOLLOW-UP OBSERVATIONS}\label{sec:obs}

\begin{table}
\centering 
\caption{General information about GPX-1. Position, distance and motion information are based on Gaia DR2 data \citep{2018AJ....156...58B}. $B$, $V$, ${g^\prime}$, ${r^\prime}$ and ${i^\prime}$ magnitudes are from the APASS catalog \citep{2016yCat.2336....0H}, $J$, $H$, $K_{\mathrm{s}}$ magnitudes -- from the 2MASS catalog \citep{2003yCat.2246....0C}, $W_1$, $W_2$, $W_3$ -- from the {\em WISE} catalog \citep{2012yCat.2311....0C}.}
\begin{tabular}{ll}
\hline
Identifiers & GPX-1\,b\\
& GPX-TF8A-1859 (GPX input catalog ID)\\
& 2MASS 02332859+5601325\\
& Gaia DR2 457317534880081152\\
& 1SWASP J023328.42+560133.4\\
& GSC 3691:0475\\
& TESS TIC~245392284\\
RA (J2000.0) & 02h 33m 28.606s\\ 
DEC (J2000.0) & +56$^{\circ}$ 01${^\prime}$ 32.55${^{\prime\prime}}$\\
Gal $l$ & 136.893301$^{\circ}$\\
Gal $b$ & -4.055172$^{\circ}$\\
Distance & $655\pm17$\,pc\\
pmRA & $-3.25\pm0.067$\,mas/y\\
pmDEC & $0.05\pm0.081$\,mas/y\\
RV & $-3.7\pm5.26$\,km/s\\
TESS \,mag & $11.90\pm0.01$\\
APASS $B$\,mag & $12.74\pm0.08$\\
APASS $V$\,mag & $12.27\pm0.05$\\
APASS ${g^\prime}$\,mag & $12.46\pm0.07$\\
APASS ${r^\prime}$\,mag & $12.18\pm0.04$\\
APASS ${i^\prime}$\,mag & $12.12\pm0.03$\\
2MASS $J$\,mag & $11.35\pm0.026$\\
2MASS $H$\,mag & $11.22\pm0.024$\\
2MASS $K_{\mathrm{s}}$\,mag & $11.18\pm0.021$\\
{\em WISE} $W_1$\,mag & $11.08\pm0.025$\\
{\em WISE} $W_2$\,mag & $11.10\pm0.022$\\
{\em WISE} $W_3$\,mag & $10.64\pm0.077$\\
\hline
\end{tabular}
\label{tab:coordinates}
\end{table}

\subsection{GPX transit detection photometry}

\begin{figure*}
  \includegraphics[width=\textwidth]{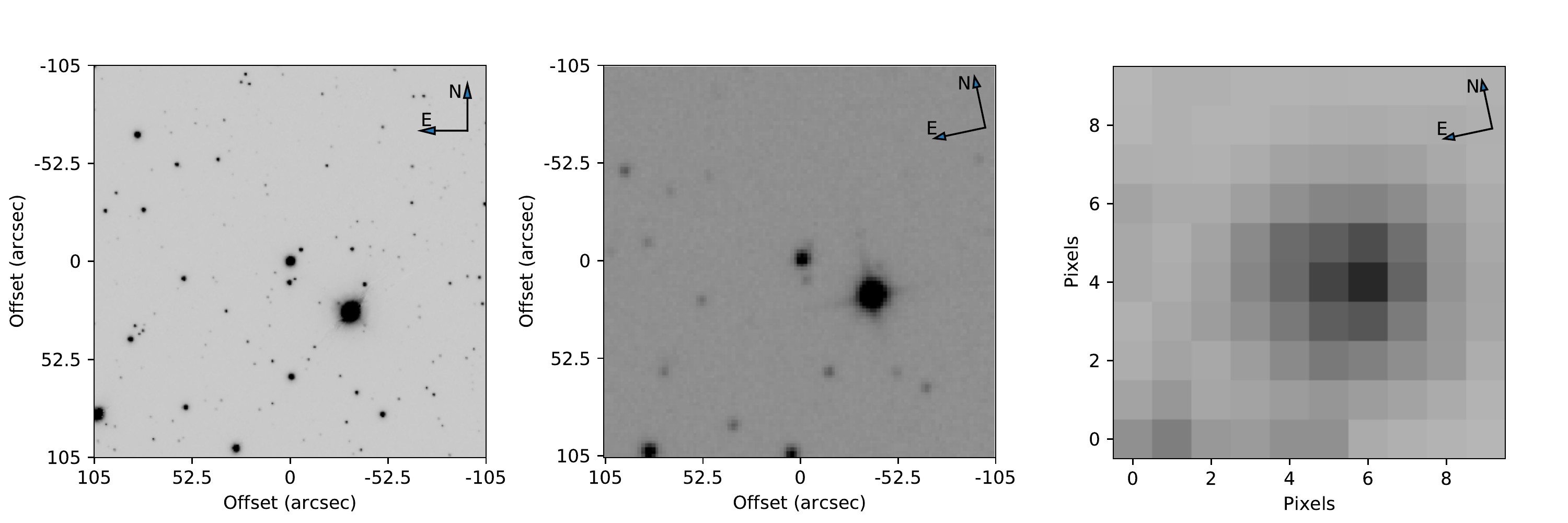}
  \caption{Left: Pan-Starrs $i'$ image of a $210\,\times\,210\,\mathrm{arcsec^2}$ region around GPX-1 ($V\sim$12\,mag), obtained with an image scale of 0.25\,arcsec\,pixel$^{-1}$. Note the bright star HD\,15691 ($V\sim$9\,mag) located 42\,arcsec SW from GPX-1. Middle: image of the same field obtained with a telescope with an image scale of 1.85\,arcsec\,pixel$^{-1}$. Right: TESS $210\,\times\,210\,\mathrm{arcsec^2}$ ($10~\times~10~\mathrm{pixel^2}$) image of the same Field of View (FoV).}
    \label{fig_Map_and_LC}
\end{figure*}

\begin{figure}
\includegraphics[width=\columnwidth] {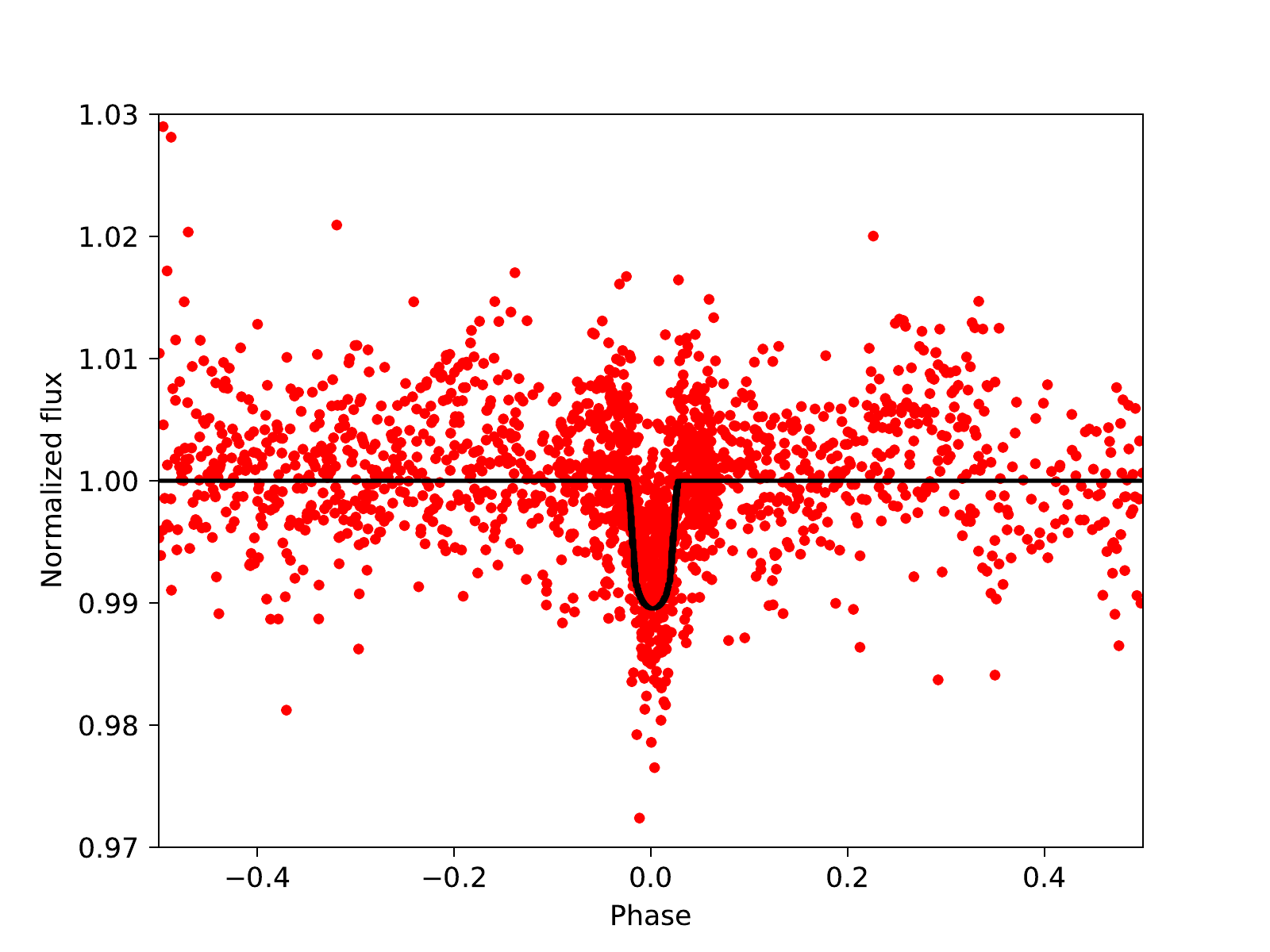}
 \caption{GPX discovery light curve as obtained with the RASA 11" wide-field telescope and folded with $\sim$1.75\,d period. The solid black line represents the best-fit transit model.}
    \label{fig_LC}
\end{figure}

The GPX survey telescope was built from readily available commercial equipment: a RASA 11" wide-field telescope (279\,mm, f/2.2; Celestron, Torrance, CA USA), based at the private observatory Acton Sky Portal in Acton, MA, USA. The GPX survey covered 10 star fields with a Field of View (FoV) of $1.67\times1.26$~deg$^2$ and 54 fields with a $2.52\times2.02$~deg$^2$ FoV in 2015-2019, which were observed for 3-5 months with 150-180 hours of image data collected per field. A detailed description of the GPX survey observations is presented in \citet{2018PASP..130g4401B}. 

The star GPX-1 (see Table~\ref{tab:coordinates}) was observed with the RASA telescope from September to December 2016 with the Johnson-Cousins \textit{$R_\mathrm{c}$} band. Survey images were treated with the K-pipe pipeline described in \citet{2016MNRAS.461.3854B}. In short, the pipeline performs a calibration of the FITS images (bias, dark and flat-field corrections), extracts fluxes of all the stars in the FoV, conducts differential photometry, and performs a period search with the Box-fitting Least Squares method (BLS; \citealt{2002A&A...391..369K}). Stars with low-noise light curves were selected as high-priority and their phase-folded light curves were examined visually.  Using BLS, a strong  peak corresponding to a $\sim$1.7\,d period in the periodogram of GPX-1 was found. The phase-folded light curve of GPX-1 is presented in Fig.~\ref{fig_LC}. After obtaining an initial photometric ephemeris, we performed photometric, speckle-interferometric, and spectroscopic follow-up observations to investigate the nature of the system.

\subsection{Photometric follow-up observations}

We conducted a photometric follow-up campaign to observe transits of GPX-1\,b with several goals: to confirm that the signals found in the GPX wide-field data were real; to refine the transit ephemeris; and to compare the depths of the transits in different filters to check their chromaticity to verify that the signal was not caused by an eclipsing binary. A dozen different telescopes participated in the follow-up campaign. The majority of those observations were performed by the small- and middle-aperture telescopes, which participate in the EXPANSION project (EXoPlanetary trANsit Search with an International Observational Network; \citealt{2018MNRAS.480..291S}). In particular, multi-colour observations of two GPX-1\,b transits in 2017 with the 1-m telescope T100 of the T\"UB\.{I}TAK National Observatory of Turkey (TUG) helped us to discard evident eclipsing binary scenario and allowed to request spectroscopic observations with the SOPHIE spectrograph (see sub-section~\ref{sophie_spec}). We used all available data for the initial characterisation of the system, but here we will present only the most precise light curves from our observational sets, which were used to derive the system parameters.

High-precision light curves of GPX-1 were obtained with four telescopes (see Table~\ref{tab:TblObs}). Five transits were observed with the RC600 telescope of the Caucasian Mountain Observatory (CMO) of Sternberg Astronomical Institute (SAI) of Moscow State University (MSU) in $g'$, $r'$, and \textit{$R_\mathrm{c}$} filters \citep{Berdnikov2020}. Three partial transits were observed with the TRAPPIST-North telescope \citep{2017JPhCS.869a2073B,2011Msngr.145....2J,2011EPJWC..1106002G} at the Ouka\"imeden Observatory in $z'$ filter and in a custom near-infrared filter $I+z'$ (transmittance $>90\%$ from 700\,nm). One full transit in \textit{$R_\mathrm{c}$} and \textit{$B$} bands were obtained with the AZT-11 telescope of the Crimean Astrophysical Observatory (CrAO) and 1-m telescope of the Deokheung Optical Astronomy Observatory (DOAO) respectively. The MASTER telescope at the Kourovka Observatory was used to observe GPX-1 in out-of-transit phase in $B$, $V$, $R$, $I$ filters to construct a Spectral Energy Distribution (SED). 

\begin{table*}
\centering 
\caption{Photometric follow-up observations log.}
\begin{tabular}{llllll}
\hline
Observatory & Telescope  & Detector & Date (filter)\\
\hline
CMO SAI MSU & RC600, D=60\,cm, F/7 & Andor iKon-L BV & 15 Sept 2019 ($r'$), 13 Oct 2019 ($r'$) &\\
\phantom{text} & \phantom{text} &\phantom{text} & 8 Nov 2019 (\textit{$R_\mathrm{c}$}), 22 Nov 2019 ($r'$)\\
\phantom{text} & \phantom{text} &\phantom{text} &  06 Dec 2019 ($g'$)\\
Ouka\"imeden & TRAPPIST-N, D=60\,cm, F/8 & Andor iKon-L BEX2 DD & 16 Aug 2019 ($z'$), 23 Aug 2019 ($z'$) \\
\phantom{text} & \phantom{text} &\phantom{text} & 13 Sept 2019 ($I+z'$)\\
CrAO & AZT-11, D=125\,cm, F/13 & FLI ProLine PL230 & 8 Nov 2019 (\textit{$R_\mathrm{c}$}) \\
DOAO & 1-m, D=100\,cm, F/8 & FLI ProLine PL-16803 & 21 Jan 2019 (\textit{$B$}) \\
\hline
\end{tabular}
\label{tab:TblObs}
\end{table*}

\begin{figure}
\includegraphics[width=\columnwidth]{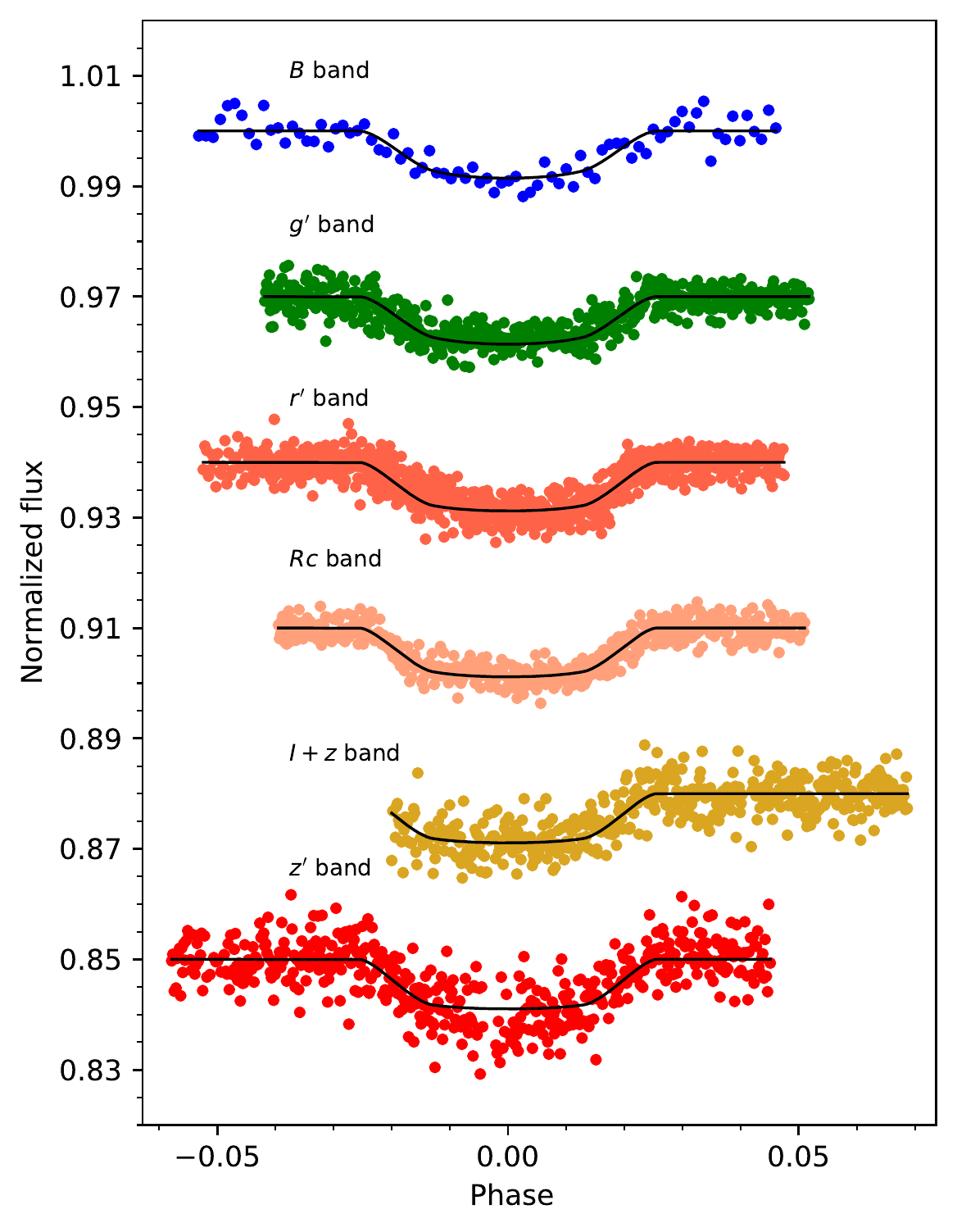}
\caption{Transits of GPX-1\,b in different bands as obtained with RC600, TRAPPIST-North, AZT-11 and DOAO 1-m telescopes. The observations are phase-folded, multiplied by the baseline (2nd order polynomial), and are shown in different colours (depending on the filter used). For each transit, the solid black line represents the best-fit model. All transits were normalised and shifted vertically for visual clarity.} 
\label{precise_lcs}
\end{figure}

All obtained images were treated similarly: data reduction consisted of dark and flat-field corrections and aperture photometry with IRAF/DAOPHOT \citep{Tody1986}. Photometric extraction was performed with different aperture sizes which were defined as $0.8, 1.0, 1.2, 1.5, 2.0, 3\,\times\,$FWHM, where FWHM is the mean full width at half maximum of of the stars' point spread function (PSF) in the image. Selection of the best aperture and comparison stars was made based on the minimization of the out-of-transit scatter of the light curve. We also controlled that the aperture centered around GPX-1 did not include any nearby stars. The resulting light curves are presented in Fig.~\ref{precise_lcs}, after period-folding and being multiplied by baseline polynomials to correct for systematic errors (see sub-section~\ref{subsec:global_an} for details). Achromatic transits with depths 0.9\% are clearly visible, which strengthens the hypothesis that the transiting body is not a star.

For the SED fit, we measured instrumental magnitudes of GPX-1 and nearby stars in the FoV and then used the APASS catalog \citep{2016yCat.2336....0H} to convert instrumental magnitudes to the standard ones.

\subsection{TESS and WASP photometry}

GPX-1 (TIC~245392284) was observed at 30-min integrations by TESS in Sector 18 in November 2019, which spans 13 orbital periods of GPX-1\,b. Since GPX-1 was not observed with 2-min-cadence, the Science Processing Operations Center pipeline (SPOC; \citealt{2016SPIE.9913E..3EJ}) did not extract photometric flux nor attempted a transit search. The light curve of GPX-1 was extracted by the Quick-Look Pipeline (QLP; \citealt{2018ApJ...868L..39H}, Huang et al. 2020 in prep.), but it was not included for candidate vetting procedures as in the vast majority of cases vetting is carried out for exoplanet candidates brighter than TESS\,mag of 10.5 (and TESS\,mag of GPX-1 is 11.9). However, even if GPX-1 light curve is included for vetting, it does not pass a predefined standard threshold for a transit signal (C.~X.~Huang, personal communication, 3 September 2019).

%The likely reason for a non-detection by the QLP is the blending of GPX-1 with the aforementioned nearby 3.4\,mag brighter star HD 15691 42\,arcsec away (see left and middle panels of Fig.~\ref{fig_Map_and_LC}).

\begin{figure}
\includegraphics[width=\columnwidth]{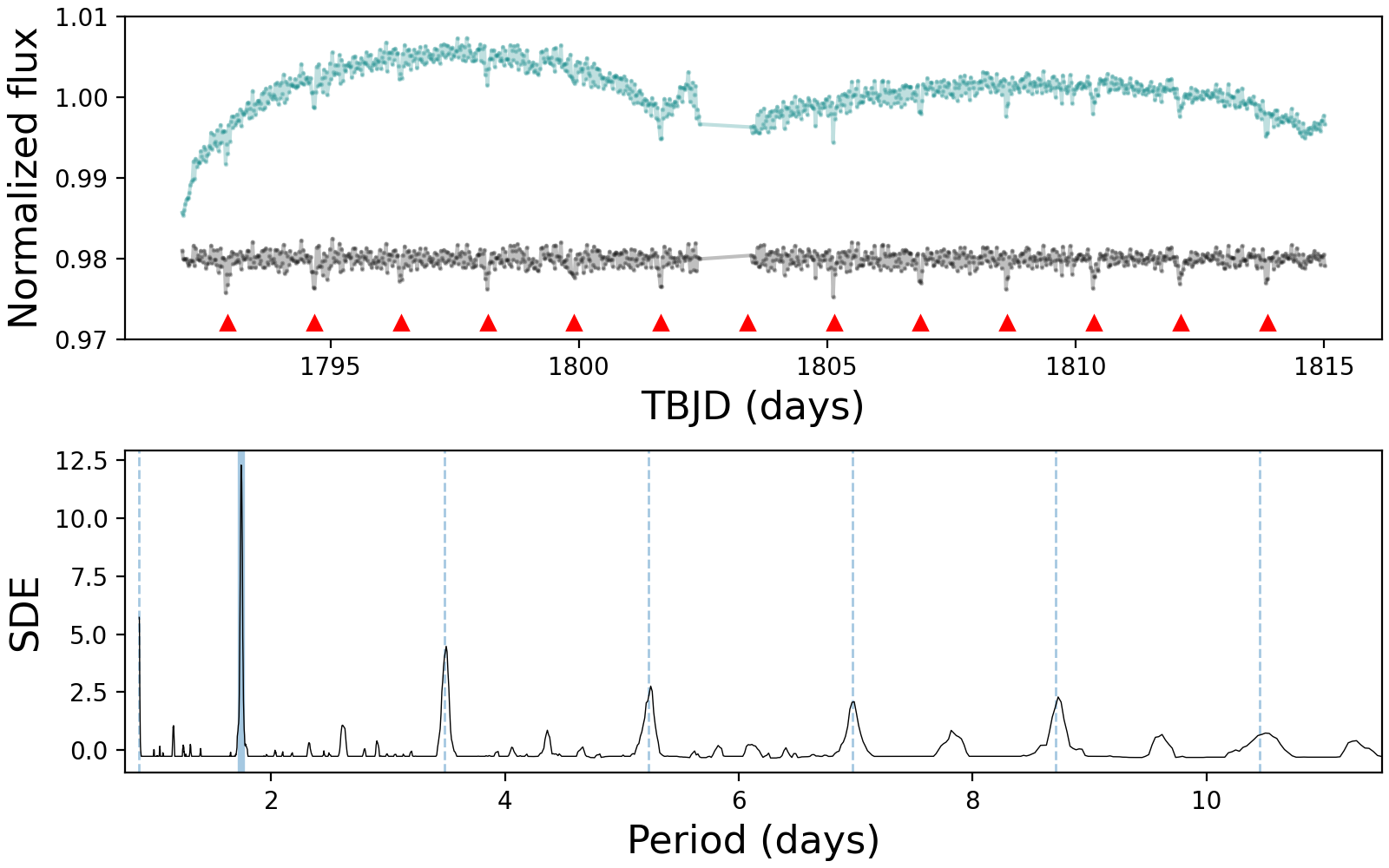}
\caption{Upper panel: TESS data for GPX-1 (TIC~245392284) from Sector 18. The teal line corresponds to the light curve from the Full Frame Images using a 2-pixel aperture, and black line
is the final clean light curve. The red triangles mark the $\sim$1.74\,d period signal. Bottom panel: periodogram yielded by the transit search performed with the
SHERLOCK pipeline, where the main $\sim$1.74\,d period signal and its harmonics are highlighted.} 
\label{fig:search}
\end{figure}

\begin{figure}
\includegraphics[width=\columnwidth]{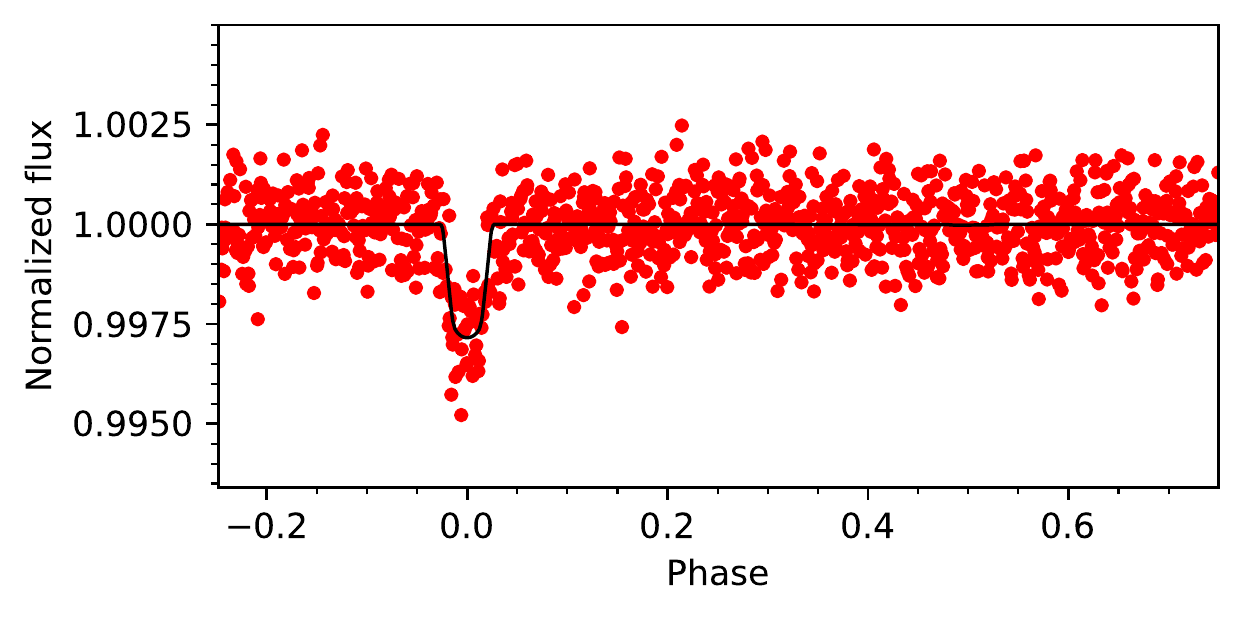}
\caption{Phase-folded light curve of GPX-1\,b as obtained using the TESS data. Due to a blending with a nearby star, obtained transit depth is $\sim$0.25\,\% (compare to $\sim$0.9\,\% from the ground-based data. See Fig.~\ref{precise_lcs}). The solid black line represents the best-fit transit model.} 
\label{fig:tess_lc}
\end{figure}

We obtained the light curve from the TESS Full Frame Images (FFIs) using a 2-pixel aperture. The raw light curve (not corrected for systematics) is presented in the upper panel of Fig.~\ref{fig:search}. We removed the systematic trends in the data by performing decorrelation with quaternions (high-cadence vector time-series that describe TESS attitude based on observations of guide stars; \citealt{2019ApJ...881L..19V}) and the background flux outside the aperture.  Then, we made use of the SHERLOCK pipeline \citep{pozuelos2020} to search for transit signals in the data. The pipeline performed transit search by means of the Transit Least Squares package (TLS; \citealt{2019A&A...623A..39H}). We successfully recovered the $\sim$1.7\,d period signal with a signal detection efficiency (SDE) of 12.3 and signal-to-noise ratio (SNR) of 18.0 (see Fig.~\ref{fig:search}). The SHERLOCK pipeline keeps searching for transit signals until there is no other signal left with SDE\,>\,5 and SNR\,>\,5 in the data. In our case, no other signals were found. Though the TESS light curve is heavily blended with the nearby 3.4\,mag brighter star HD\,15691 (see Fig.~\ref{fig_Map_and_LC}), we included it in our global modelling of the system with a dilution term, i.e. allowing the transit depth to be set by the ground-based observations. The resulting period-folded light curve is presented in Fig.~\ref{fig:tess_lc}, where a $\sim$0.25\,\% transit is visible. We did not detect any signs of occultation signal in the TESS data and we place a 3-$\sigma$ upper limit on occultation depth of 0.04\,\%, what corresponds to $\sim$0.1\% if there was no dilution from the nearby star in the TESS data. This limit translates into an upper limit on BD's day side effective temperature of 3800\,K, which does not contradict the equilibrium temperature from the global solution (see Section~\ref{sec:discuss}).

GPX-1 was also observed by the WASP-North telescope, which is now decommissioned \citep{Pollacco2006}. Observations were obtained in 2004-2008 (WASP ID 1SWASP~J023328.42+560133.4\footnote{WASP time-series viewer is available at \url{https://exoplanetarchive.ipac.caltech.edu}}). The WASP telescope had a better image scale than TESS, but blending with aforementioned nearby star HD\,15691 coupled with a poorer photometric precision (compared to TESS), prevented robust transit detections in the data. 

\subsection{Speckle-interferometric observations}

\begin{figure}
\centering
\includegraphics[width=1\columnwidth]{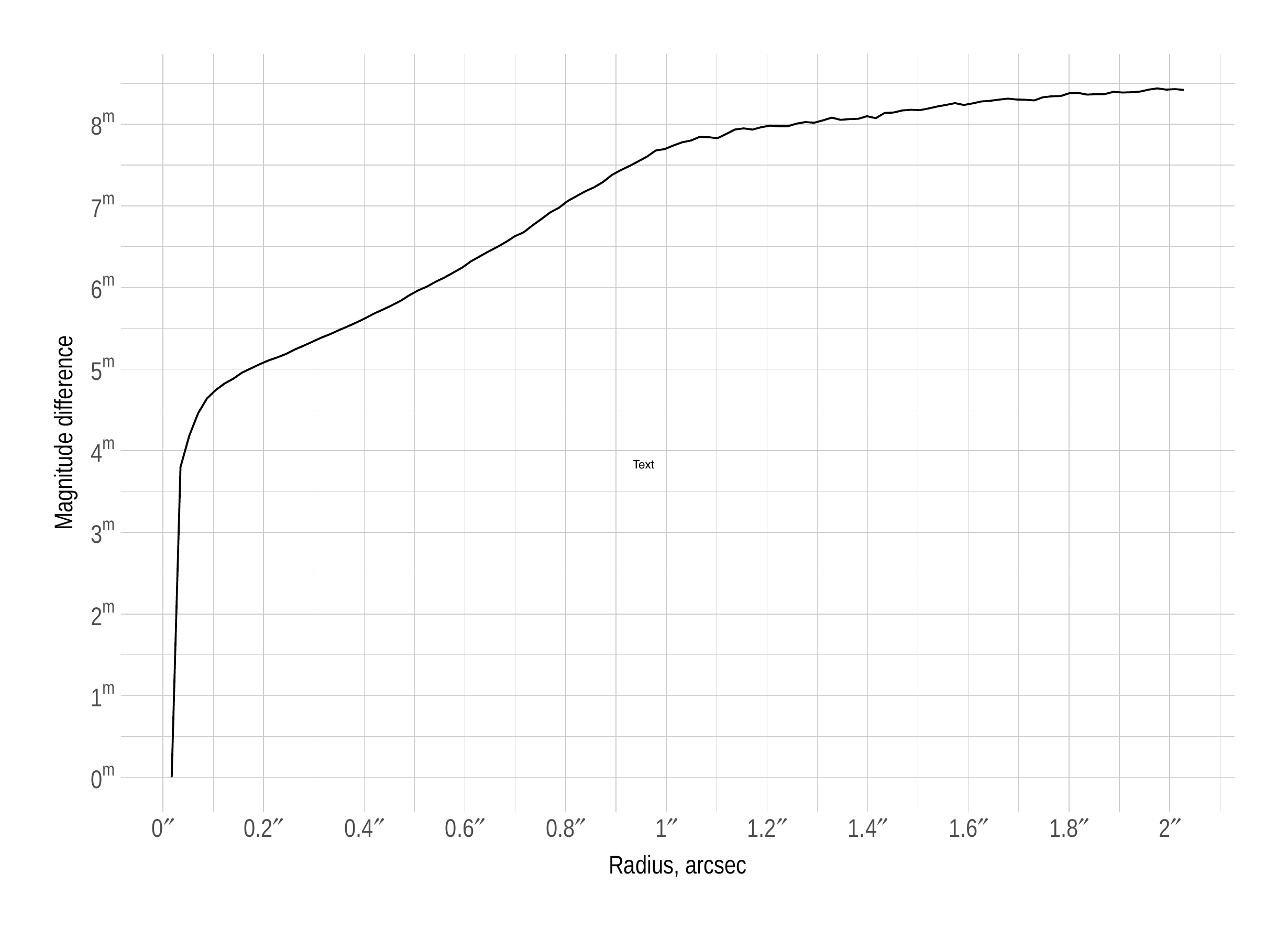}
\caption{Results of SAO speckle-interferometer observations of GPX-1. The 3-$\sigma$ detection limit is plotted as a function of radius. The limits are 0.03\,arcsec for a brightness difference $\Delta m$ = 0\,mag, 0.05\,arcsec for $\Delta m$ = 3\,mag and 0.1\,arcsec for $\Delta m$ = 4.5\,mag.}
\label{speckle_limits}
\end{figure}

We carried out speckle-interferometric observations of GPX-1 to identify any nearby companions, which might affect our transit photometry and spectroscopy, or potentially be the true source of the transit signal. Observations were performed with the use of the 6-m telescope of the Special Astrophysical Observatory (SAO) of the Russian Academy of Sciences (RAS) on 8 October and 1 December 2017. Both observations were performed using SAO speckle-interferometer \citep{2009AstBu..64..296M} with $4.5 \times 4.5$\,$\mathrm{arcsec^2}$ FoV with a 100\,nm-wide filter centred on the wavelength of 800\,nm, which corresponds to the photometric $I$ band. The length of the series was 2000 frames, and the exposure time per frame was 100\,ms and 50\,ms for the nights of 8 October and 1 December 2017, respectively. Image processing included the calculation of the average power spectra of the series and construction of the corresponding auto-correlation functions. The search for companions was done by analysing auto-correlation functions. The calculated limits for the presence of components near GPX-1 are shown in Fig.~\ref{speckle_limits}. Based on the analysis of the SNR by series, we were unable to detect a visual companion of GPX-1, and we provide conservative limits on the presence of a secondary companion. The limits are 0.03\,arcsec for a brightness difference $\Delta m$ = 0\,mag, 0.05\,arcsec for $\Delta m$ = 3\,mag and 0.1\,arcsec for $\Delta m$ = 4.5\,mag. These limits are consistent with the data obtained in both observational sets.

\subsection{Spectroscopic observations}{\label{sophie_spec}}

GPX-1 was observed with the SOPHIE spectrograph to obtain RV data to measure the mass of the transiting body. SOPHIE is dedicated to high-precision RV measurements
with the 1.93-m telescope of the Haute-Provence Observatory \citep{2008SPIE.7014E..0JP,2009A&A...505..853B}. We used SOPHIE in High-Efficiency
mode with a resolving power $R=40\,000$ and slow readout mode. We obtained 16 observations between October 2017 and March 2018. Depending on observing
conditions, exposure times ranged between 9 and 32~min in order to maintain an SNR of $\sim$20 per pixel as constant as possible throughout the observations. This SNR was chosen as a compromise between precision and exposure time.

The spectra were extracted using the standard SOPHIE data reduction pipeline. We used a weighted cross-correlation function (CCF) with a G2-type numerical mask \citep{1996A&AS..119..373B,2002A&A...388..632P}. We also tried other numerical masks and they provide results in agreement with the G2-type mask, but with larger uncertainties. The CCF has contrasts representing only $\sim$2.2\,\%\ of the continuum and is broad, with a full width at half maximum (FWHM) of $64 \pm 3$\,km/s. 

\begin{figure}
\includegraphics[width=1.0\columnwidth]{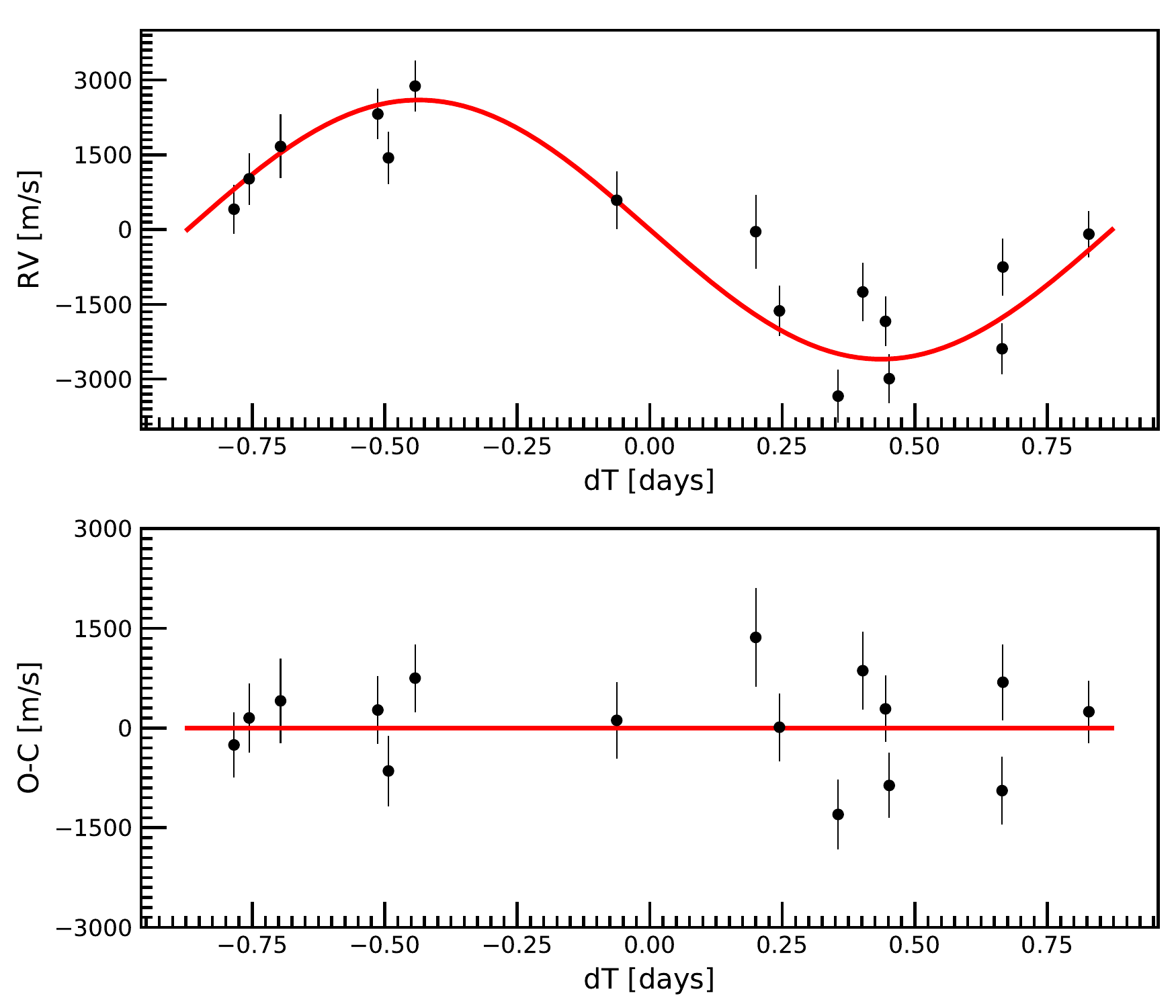}
\caption{Top: SOPHIE radial velocity (RV) measurements of GPX-1 phase-folded at the 1.75~days period of the BD with the imposed best-fit Keplerian model. Bottom: best-fit residuals.} 
\label{fig:RVs}
\end{figure}

\begin{figure}
 \centering
\includegraphics[width=1.0\columnwidth, trim={2cm 2cm 0cm 2cm}]{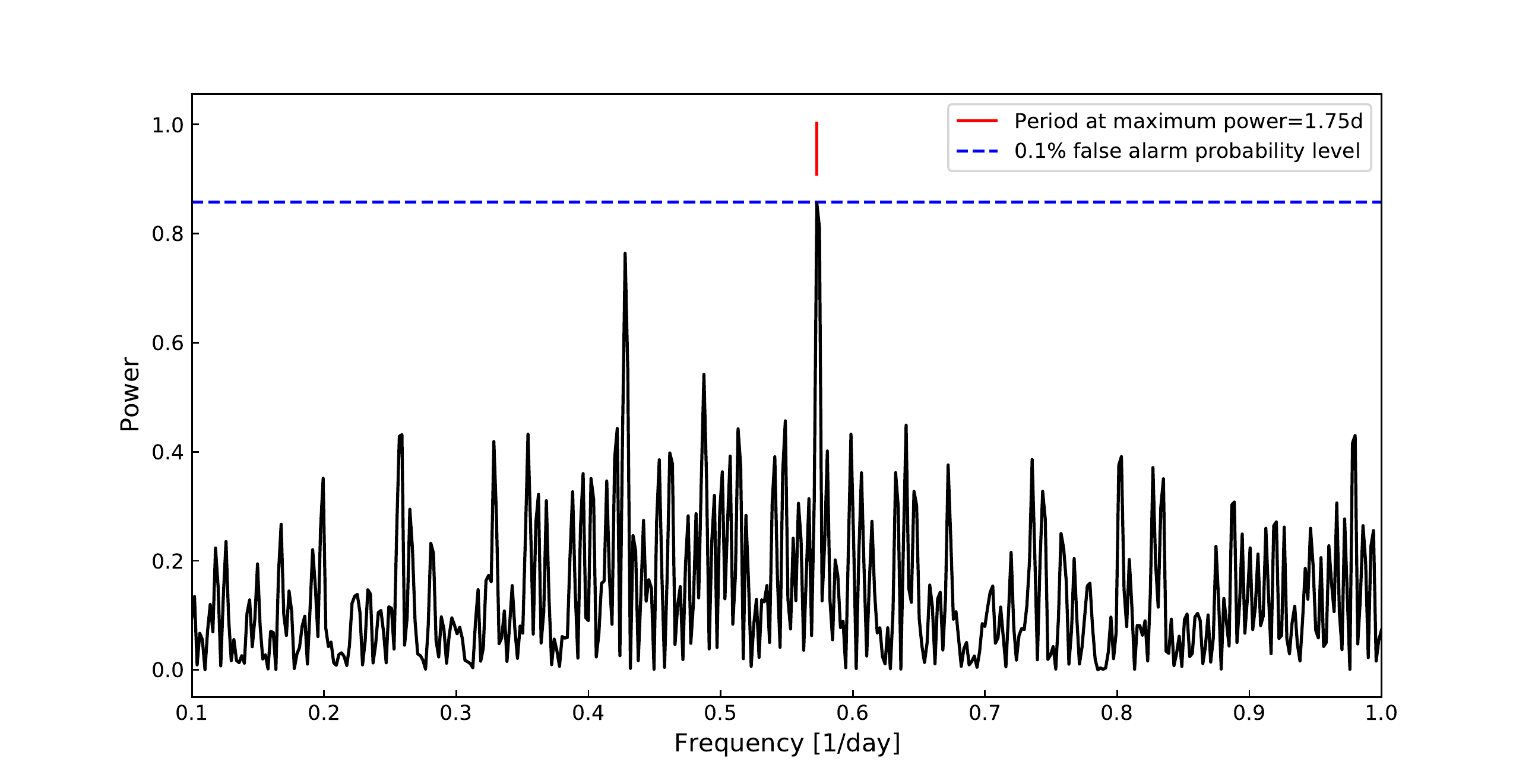}
\caption{The Lomb-Scargle periodogram of the SOPHIE radial velocity (RV) measurements. The most powerful peak on the periodogram corresponds to a period of 1.75\,d.}
  \label{RV_periodogram}
\end{figure}

We fitted the CCFs with Gaussians to measure the RVs, their errors bars, and the bisector spans (Table~\ref{table_rv}). Such broad and shallow CCFs allowed us to determine RVs with only a relatively low precision of $\pm 500$\,m/s. Still, the RVs show significant variations, which are in agreement with the period and phase obtained from the photometric transits (see Fig.~\ref{fig:RVs} with the RV measurements and Fig.~\ref{RV_periodogram} with the Lomb-Scargle periodogram of the RVs). The semi-amplitude of a simple fit to the RVs is of order 2.5\,km/s which would correspond to a $\sim$20\,$M_{\mathrm{Jup}}$ companion.

\begin{figure}
 \centering
\includegraphics[width=1.0\columnwidth]{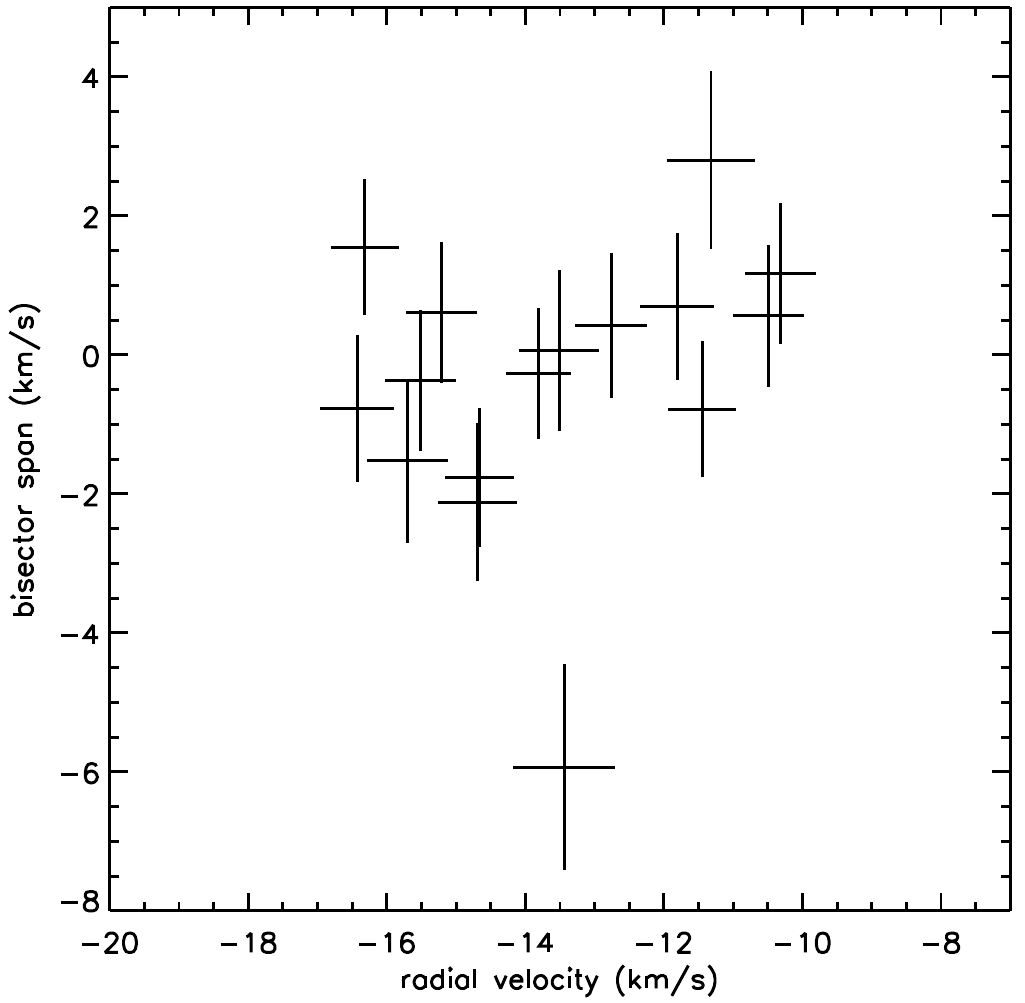}
\caption{GPX-1 bisector span as a function of the radial velocities (RVs) with 1-$\sigma$\,error bars.}
  \label{figure_biss}
\end{figure}

To check if RV variations were caused by a blending scenario of stars with different spectral types, we measured RVs with a set of different stellar masks. In all cases, we obtained RV variations with similar amplitudes and we can conclude that the blending scenario is very unlikely. To assess possible change of the shape of spectral lines due to stellar activity or blends, we measured CCF bisector spans. Our measurements have low precision due to the broad, shallow CCFs, but they show no correlations with the RV variations (a Spearman correlation coefficient is 0.33; see Fig.~\ref{figure_biss}). Also, the visual inspection of the CCF at a large velocity span does not suggest the presence of any background target. Therefore, we can conclude that the observed RV variations are caused by a sub-stellar companion.

\begin{figure}
 \centering
\includegraphics[width=1.0\columnwidth, trim={2cm 2cm 2.5cm 2cm}]{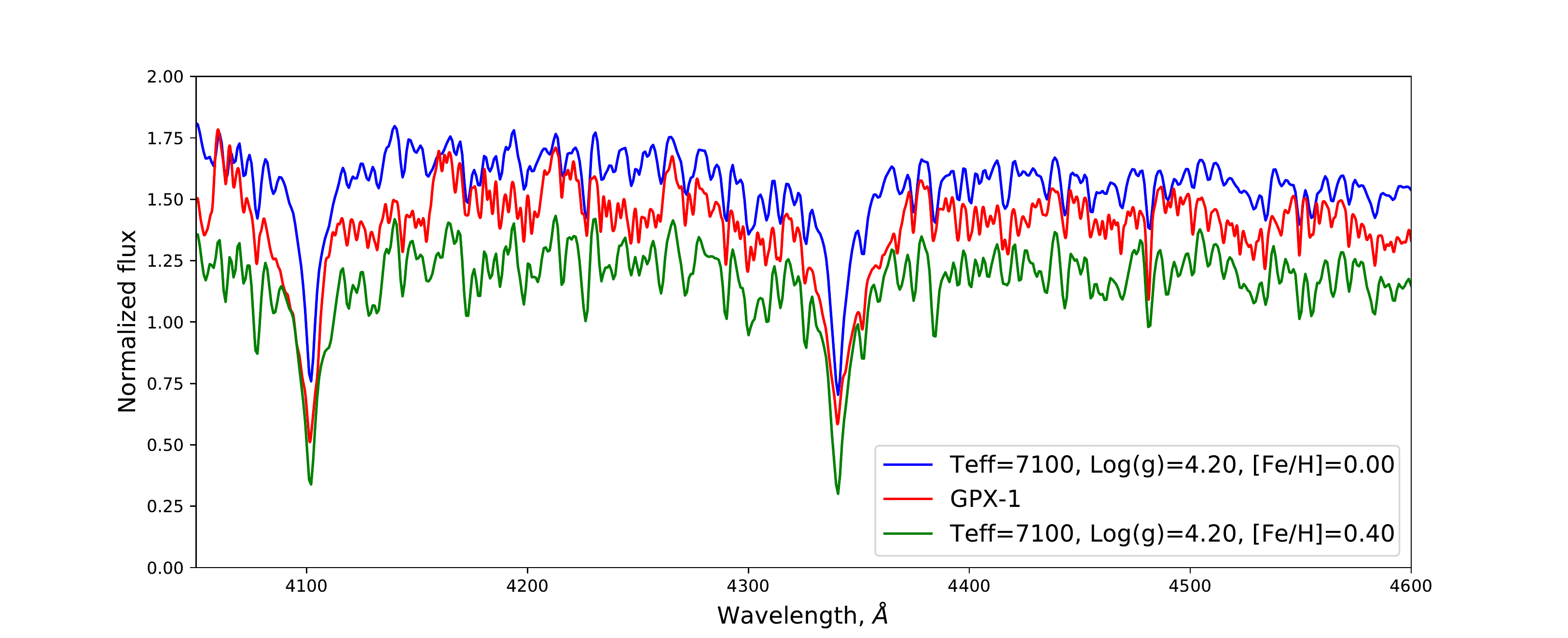}
\caption{Filtered spectrum of the GPX-1 star in the 4000-4600\,{\AA} range and templates from the MILES spectral library \citep{2011A&A...532A..95F}. All spectra were normalized and shifted vertically for visual clarity.}
  \label{spec_templates}
\end{figure}

We carried out spectral classification and determination of metallicity using filtered and smoothed SOPHIE spectrum. First, we a applied a median filter with a kernel of 0.2\,{\AA} for suppression of short spikes from hot pixels, cosmic rays, etc. Next, we applied a Gaussian filter with a kernel of 0.4\,{\AA} to enable matching of our spectral resolution with the templates from spectral libraries. We used a strength of hydrogen lines as a temperature criteria, as they are the least affected by differences in metallicity, and the G-band, which starts to appear in early F-type stars and which is slightly evident in our filtered spectrum. At the next step, we compared the object spectrum with templates from the CFLIB  \citep{2011A&A...525A..71W} and the MILES \citep{2011A&A...532A..95F} spectral libraries. Parameters of the best matched templates were used to estimate the effective temperature, surface gravity and metallicity (see Fig.~\ref{spec_templates}). We classified GPX-1 as an F2 star and we found stellar parameters to be: $T_{\mathrm{eff}}=7100\pm150$\,K, log$\,g_{\star}=4.1\pm0.1$, $\lbrack \mathrm{Fe/H}\rbrack=0.2\pm0.1$. We provide conservative error bars as our spectrum has a relatively low SNR.

We could evaluate the line broadening parameter $v\sin{\,i_*}=40\pm10$\,km/s derived from the measured widths of the CCFs using the calibration of \citet{2010A&A...523A..88B}. This calibration stands for slower rotators and we evaluated a conservative uncertainty, but our estimation should be taken with caution. Thus, GPX-1 will benefit from additional spectroscopic observations.

\begin{table}
\caption{SOPHIE measurements of GPX-1.}
\begin{center}
\begin{tabular}{cccrrr}
\hline
BJD$_{\rm UTC}$ & RV & 1-$\sigma$ & bisect.$^\ast$ & exp. & SNR$^\dagger$ \\
-2\,458\,000 & (km/s) & (km/s) & (km/s)  & (sec) &  \\
\hline
038.5049  &  -12.76  &  0.52  &  0.42  &  792    &  23.0  \\
041.4569  &  -16.32  &  0.49  &  1.55  &  536    &  23.5  \\       
053.5725  &  -16.43  &  0.53  &  -0.77 &  795    &  22.7  \\       
054.4488  &  -10.49  &  0.51  &  0.56  &  978    &  21.9  \\       
055.6264  &  -15.21  &  0.51  &  0.61  &  1207   &  22.4  \\       
057.5347  &  -13.81  &  0.47  &  -0.27 &  723    &  22.9  \\       
083.5412  &  -14.69  &  0.57  &  -2.12  &  1800   &  18.0  \\     
084.5573  &  -13.51  &  0.58  &  0.06  &  1800   &  19.3  \\       
085.5799  &  -11.45  &  0.49  &  -0.78  &  844    &  23.1  \\       
121.4556  &  -13.44  &  0.74  &  -5.93  &  1004   &  19.7  \\       
123.4021  &  -15.70  &  0.59  &  -1.52  &  875    &  22.3  \\       
171.3555  &  -11.81  &  0.53  &  0.70  &  1362   &  23.0  \\       
172.2931  &  -14.66  &  0.50  &  -1.77  &  1307   &  22.0  \\     
177.3268  &  -15.51  &  0.51  &  -0.37  &  1919   &  23.4  \\       
204.2990  &  -11.32  &  0.64  &  2.80  &  1800   &  18.6  \\       
206.2974  &  -10.32  &  0.51  &  1.17  &  1100   &  23.0  \\ 
\hline
\multicolumn{6}{l}{$\ast$: bisector spans; error bars are twice those of the RVs.} \\
\multicolumn{6}{l}{$^\dagger$: SNR per pixel at 550\,nm.} \\
\end{tabular}
\end{center}
\label{table_rv}
\end{table}

\section{Data Analysis}\label{sec:analysis}

\subsection{General considerations}

GPX-1 has a spectral type F2 and is a fast rotator with a line broadening parameter $v\sin{\,i_*}=40\pm10$\,km/s. This combination makes the inference of stellar parameters, such as the effective temperature $T_{\mathrm{eff}}$, metallicity \lbrack Fe/H\rbrack, and surface gravity log$\,g_{\star}$ using spectra with low SNR challenging and unreliable due to the relatively small number of spectral absorption lines and their broadening. However, knowledge of a host star is crucial as all the parameters of a companion are obtained relative to the host star, primary $R_{\star}$ and $T_{\mathrm{eff}}$. 

To address this problem, we used the MCMCI tool to perform an integrated modelling of GPX-1 and its transiting BD. MCMCI is a code that utilises Markov Chain Monte Carlo (MCMC) stochastic simulations of the RV and photometric data to analyse the posterior probability distribution function (PDF) of the model parameters coupled with the isochrone placement algorithm. The MCMCI code and the logic behind it are described in details in \citet{2020A&A...635A...6B} and here we outline the main aspects of our analysis. The MCMCI tool performs an integrated three-step analysis:
\begin{enumerate}
    \item Photometric light curves are analysed to infer a mean stellar density $\rho_{\star}$;
    \item The stellar density $\rho_{\star}$ together with the stellar metallicity \lbrack Fe/H\rbrack, and effective temperature $T_{\mathrm{eff}}$ (computed from photometric colour) are used to infer the stellar mass $M_{\star}$, radius $R_{\star}$ and age by the isochrone placement algorithm, which considers pre-computed grids of isochrones and tracks based on the PAdova and TRieste Stellar Evolutionary Code (PARSEC, v1.2S; \citealt{2017ApJ...835...77M}). The isochrone placement algorithm is presented in detail in \citet{2015A&A...575A..18B,2016A&A...585A...5B};
    \item Inferred stellar parameters are used to recover companion's radius $R_{\mathrm{BD}}$ and mass $M_{\mathrm{BD}}$ using the photometric and RV data sets.
\end{enumerate}

Similar to the MCMC code described in \citet{2012A&A...542A...4G}, the MCMCI tool utilises a Keplerian model by \citet{2010exop.book...15M} and a transit model by \citet{2002ApJ...580L.171M} to jointly fit the RV and photometric data.

\begin{figure}
  \includegraphics[width=1\columnwidth]{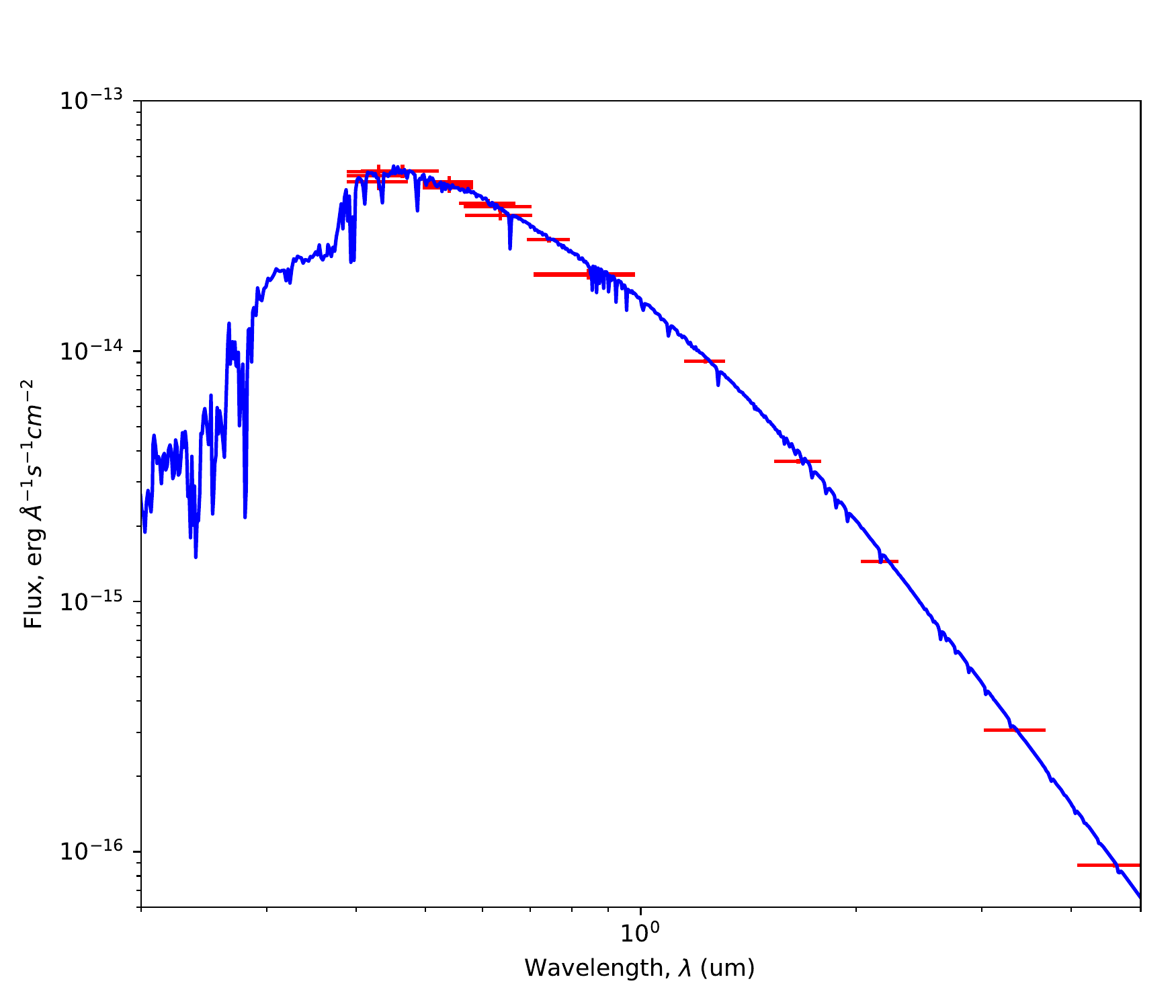}
  \caption{The APASS ($B$, $V$, \textit{g${^\prime}$, r${^\prime}$, i${^\prime}$}), 2MASS ($J, H$, and $K_{\mathrm{s}}$), {\em WISE} ($W1$, $W2$), MASTER ($B$, $V$, $R$, $I$) and TRAPPIST-North ($B$, $V$, $R_\mathrm{c}$, $I_\mathrm{c}$) spectral points are shown. 
  %Some markers are intersected. 
  The blue curve is a model for the star with $T_{\mathrm{eff}}=7200$\,K, $R_{\star}=1.64\,R_{\odot}$, log$\,g_{\star}=4.2$ and $\lbrack \mathrm{Fe/H}\rbrack=0.43$ at 655\,pc and $E(B-V)=0.18$.}
    \label{SED}
\end{figure}

We would like to note that interstellar dust extinction plays a major role when observing a distant object with low galactic latitudes by attenuating and reddening observed photometric magnitudes. In the case of GPX-1, its distance is 655 $\pm$ 17\,pc \citep{2018AJ....156...58B} and its galactic latitude is $b\sim$~-4\,deg which suggests non-negligible effect of dust extinction. Therefore, as we compute effective temperature $T_{\mathrm{eff}}$ from the photometric colour, we need to deredden the photometric colour index of GPX-1. To make this correction, we constructed a SED of GPX-1 using available photometric data (see Fig.~\ref{SED}). We used data from available catalogs and observations made with the MASTER telescope \citep{Gorbovskoy2013} and with the TRAPPIST-North telescope. We performed a fit using a grid of ATLAS9 stellar atmosphere models \citep{2003IAUS..210P.A20C} with a constant Gaia DR2 distance, uniform priors on log$\,g_{\star}$ in the range $3.9\,-\,4.3$ and $\lbrack \mathrm{Fe/H}\rbrack$~in the range $0.0,-\,0.4$, to determine the amount of interstellar reddening. Our choice of priors' range is justified by the analysis of the SOPHIE spectra. Our best-fit parameters are: $T_{\mathrm{eff}}=7200\pm150$\,K, $R_{\star}=1.6 \pm $0.1\,$R_{\odot}$, $\lbrack \mathrm{Fe/H}\rbrack=0.43\pm0.2$ and $E(B-V)=0.18\pm0.1$. The derived value of reddening is in a good agreement with the interpolated values from \citet{2019MNRAS.483.4277C}. Thus, GPX-1 colour and magnitude corrected for interstellar extinction are: $(B-V)_{0}=0.3\pm0.1$\,mag and $V_{0}=11.8\pm0.1$\,mag.

The stellar metallicity \lbrack Fe/H\rbrack~from the SED fit, the dereddened $(B-V)_0$ colour index, and the stellar luminosity $L_{\star}$ (computed from the dereddened $V_0$ magnitude and the distance from Gaia DR2 catalog \citep{2018A&A...616A...1G}) were used as priors. The stellar effective temperature $T_{\mathrm{eff}}$ (computed from $(B-V)_0$ colour), stellar luminosity $L_{\star}$ (computed from the $V_0$ magnitude and the distance), and mean density $\rho_{\star}$ (computed from light curves) were the input parameters for computing $M_{\star}$, $R_{\star}$, and the age from isochrones.

As the GPX-1 star is a fast rotator, its shape is oblate. This causes poles of the star to have a higher surface gravity, and thus higher temperature. We assessed the ratio between the equatorial and polar effective temperatures ${T_\mathrm{eff\_eq}} / T_\mathrm{{eff\_p}}$ as a function of flattening using the work of \citet{2011A&A...533A..43E}. We assumed that we observed GPX-1 edge on and thus $v_\mathrm{{eq}}\sim$40\,km/s. In this scenario, we obtained flattening parameter $\epsilon=1-R_\mathrm{p}/R_\mathrm{eq}<0.01$, where $R_\mathrm{p}/R_\mathrm{eq}$ is the ratio between the polar and equatorial radii. We obtained $\mathrm{T_{eff\_eq}}$ / $\mathrm{T_{eff\_p}}$ $\approx$ 1 making the effect of gravity darkening not significant. Using \citet{2009ApJ...705..683B}, we estimate that the effect of gravity darkening will not be detectable in the TESS data because of the dilution of a nearby star and because of the 30-min integration time. Thus, we did not take it into account in our modelling of the system.

\subsection{Global analysis}\label{subsec:global_an}

Before performing a global MCMC analysis of the available high-precision photometric and RV data, we performed a preliminary analysis of each light curve. The goal of this analysis was to account for correlations of the extracted photometric fluxes with the external environmental and/or instrumental parameters by obtaining a proper baseline correction model. A baseline model giving the minimum Bayesian Information Criterion (BIC, \citealt{1978AnSta...6..461S}) was selected for every light curve (see \citealt{2012A&A...542A...4G} and \citealt{2018PASP..130g4401B} for a more detailed description of the applied method). In most cases, corrections for position drift of stars on a CCD and rapid changes of FWHM were applied using 2nd order polynomial.

For our global analysis of all the data sets, the following parameters (with their normal prior distributions) were used as jump parameters in our MCMC simulations and were allowed to vary :

\begin{itemize}
% \item the luminosity of the host star $L_{\star}$ ();
  \item the brown dwarf (BD) orbital period $P$ ($1.7445\pm0.0001$\,d);
  \item the BD transit duration $W$ ($0.088\pm0.005$\,d);
  \item the BD mid-transit time $T_0$ ($2458770.24\pm0.01$\,${\mathrm{BJD_{TDB}}}$);
  \item the ratio of the BD and host star areas $(R_{\mathrm{BD}}/R_\star)^2$, where $R_{\mathrm{BD}}$ is the BD radius and $R_\star$ is the stellar radius ($0.8\pm0.2$\,\%); 
  \item the occultation depth $dF_{\mathrm{occ}}$ in TESS filter ($0.05\pm0.025$\,\%);
  \item the impact parameter $b^\prime = a\,\cos{i_{\mathrm{BD}}/R_\star}$ for a circular orbit, where $a_{\mathrm{BD}}$ is the semi-major axis and $i_{\mathrm{BD}}$ is the orbital inclination of the BD ($0.86\pm0.05$); 
  \item the parameter $K_2 = K\sqrt{1 - e^2} P^{1/3}$, where $K$ is the radial velocity orbital semi-amplitude;
  \item  $\sqrt{e}\,\sin{\omega}$ and $\sqrt{e}\,\cos{\omega}$ parameters, where $\omega$ is the argument of periastron;
  \item the effective temperature $T_{\mathrm{eff}}$ of the host star computed from $(B-V)_0$ colour and metallicity \lbrack Fe/H\rbrack~from the SED fit;
  \item the combinations $c_1 =2u_1 +u_2$ and $c_2 = u_1 - 2u_2$ of the quadratic limb-darkening coefficients $u_1$ and $u_2$.
\end{itemize}

Quadratic limb-darkening (LD) law coefficients $u_1$ and $u_2$ with normal priors distributions were interpolated from the paper by \cite{2012A&A...546A..14C} and applied to the data in each photometric band. We converted all time-stamps of our measurements from ${\mathrm{HJD_{UTC}}}$ to $\mathrm{BJD_{TDB}}$ \citep{2010PASP..122..935E}. Then, we ran one relatively short MCMC chain of 20\,000 steps to obtain a correction factor (CF) for every light curve, which was used to multiply initial photometric errors. This procedure is done to rescale the photometric errors and account for over- or underestimation of photometric noise (see \citealt{2012A&A...542A...4G} and \citealt{2018PASP..130g4401B} for the details). Once proper CFs were obtained, five 100\,000-step chains were executed with starting points inferred from perturbed BLS solutions and with a 20\% burn-in phase. Convergence of the chains was checked with the use of Gelman-Rubin statistical test \citep{1992StaSc...7..457G}. The Gelman-Rubin statistic was less than 1.04 for every jump parameter. Parameters of the BD were deduced from the set of jump parameters and stellar parameters ($M_{\star}$, $R_{\star}$) inferred from isochrones. Deduced parameters of the system for a circular orbit are presented in Table~\ref{tab:sys_params} (median values of the posterior PDFs and their respective 1-$\sigma$ limits) and the results of the analysis are presented in the next section. We also performed a second analysis of the data where eccentricity was allowed to float. We inferred a 3-$\sigma$ upper limit on orbital eccentricity of 0.1.

\begin{table}
\centering 
\caption{Inferred parameters of GPX-1 system: median values of the posterior PDFs and their respective 1-$\sigma$ limits.}
\begin{tabular}{ll}
\hline
\multicolumn{2}{c}{Output parameters from the global MCMC analysis}\\
\hline
Jump parameters & Value\\
\hline
Orbital period $P$ [d] & $1.744579\pm0.000008$\\
Transit width $W$ [d] & $0.087\pm0.002$\\
Mid-transit time $T_0$ \lbrack ${\mathrm{BJD_{TDB}}}$\rbrack & $2458770.23823\pm0.00040$\\
$b^\prime = a\,\cos{i_{\mathrm{BD}}}/R_\star$ [$R_\star$] & $0.82_{-0.03}^{+0.02}$ \\
BD/star area ratio $(R_{\mathrm{BD}}/R_\star)^2$ [\%] & $0.90\pm0.03$\\
$dF_{\mathrm{occ}}$ (TESS band) [\%] & < 0.1 (3-$\sigma$)\\
$K_2$ [m/s] & $2780\pm210$\\
$\sqrt{e}\,\sin{\omega}$ & 0 (fixed)\\
$\sqrt{e}\,\cos{\omega}$ & 0 (fixed)\\
$T_{\mathrm{eff}}$ [K] & $7000\pm200$\\
\lbrack Fe/H\rbrack$\!\!$~~[dex] & $0.35\pm0.1$ \\
$(B-V)_0$ colour [mag] & $0.37\pm0.04$ \\
Luminosity $L_{\star}$ [$L_{\astrosun}$] & $5.33\pm0.40$\\ 
\hline
Deduced stellar parameters & Value\\
\hline
Mass $M_{\star}$ [$M_{\astrosun}$] & $1.68\pm0.10$\\
Radius $R_{\star}$ [$R_{\astrosun}$] & $1.56\pm0.10$\\
Mean density $\rho_{\star}$ [$\rho_{\astrosun}$] & $0.44\pm0.10$\\
Surface gravity log$\,g_{\star}$ [cgs] & $4.27\pm0.05$\\
Age [Gyr] & $0.27_{-0.15}^{+0.09}$ \\
\hline
Deduced BD parameters & Value \\
\hline
$K$ [m/s] & $2310\pm180$\\
BD/star radius ratio $R_{\mathrm{BD}}/R_\star$ & $0.095\pm0.002$\\
Orbital semi-major axis $a$ [au] & $0.0338\pm0.0003$\\
$a/R_{\star}$ & $4.67_{-0.32}^{+0.55}$\\
Orbital eccentricity $e$ & 0 (fixed)\\
Orbital inclination $i_{\mathrm{BD}}$ [deg] & $79.9_{-0.6}^{+0.7}$\\
Argument of periastron $\omega$ [deg] & - \\
Surface gravity log$\,g_{p}$ [cgs] & $4.37\pm0.06$\\
Mean density $\rho_{p}$ [$\rho_\mathrm{Jup}$] & $6.2_{-1.0}^{+1.3}$\\
Mass $M_{\mathrm{BD}}$ [$M_\mathrm{Jup}$] & $19.7\pm1.6$ \\
Radius $R_{\mathrm{BD}}$ [$R_\mathrm{Jup}$] & $1.47\pm0.10$ \\
Roche limit $a_R$ [au] & $0.00754\pm0.00048$ \\
$a/a_{R}$ & $4.5\pm0.3$\\
Equilibrium temperature $T_{\mathrm{eq}}$ [K] & $2300\pm48$\\
Irradiation $I_{\mathrm{BD}}$ [$I_{\mathrm{Earth}}$] & $4650\pm370$\\
\hline
\end{tabular}
\label{tab:sys_params}
\end{table}

\section{Results and Discussion}\label{sec:discuss}

We obtained a mass and a radius of the host star of $1.68\pm0.10$\,$M_{\astrosun}$ and $1.56\pm0.10$\,$R_{\astrosun}$ respectively. Given an RV semi-amplitude of $2309\pm180$\,m/s and a transit depth of $0.90\pm0.03$\,\%, we estimate GPX-1\,b to have a mass of $19.7\pm 1.6$\,$M_{\mathrm{Jup}}$ and a radius of $1.47\pm0.10$\,$R_{\mathrm{Jup}}$. Based on these results from the data modelling, achromaticity of the transit depths, the lack of a visual companion near GPX-1, and a lack of bisector variations, we are confident that the results demonstrate that the F2 star GPX-1 has a BD companion.

With its $\sim$1.75\,d period and $\sim$80\,deg inclination, GPX-1\,b enters a small set of transiting and short-period BDs, where similar BDs are HATS-70\,b ($P\sim$1.9\,d; \citealt{2019AJ....157...31Z}) and KELT-1\,b ($P\sim$1.2\,d; \citealt{2012ApJ...761..123S}). All three of these BDs orbit A or F-type stars, which combined with their proximity to their host stars, leads to them receiving an extreme amount of radiation, and hence they have high equilibrium temperatures. For GPX-1\,b, its irradiation is a factor of $4650\pm370$ greater than that of the Earth, and its equilibrium temperature, assuming complete heat redistribution and zero albedo, is $2300\pm48$\,K. However, its radius is 7\% larger than that of the  $13\,M_{\mathrm{Jup}}$ HATS-70\,b ($1.384_{-0.074}^{+0.079}$\,$R_{\mathrm{Jup}}$), which receives almost twice as much irradiation as GPX-1\,b. One important difference between these two young systems is the age, with HATS-70\,b, estimated to be $0.81_{-0.33}^{+0.5}$\,Gyr and for GPX-1\,b to be only $0.27_{-0.15}^{+0.09}$\,Gyr (or $270_{-150}^{+90}$\,Myr; median value of the posterior PDF and its respective 1-$\sigma$ limits).

Unfortunately, our age estimate from the MCMCI code (which is based on the isochrone placement algorithm) has large uncertainties and GPX-1 is too hot for the application of gyrochronology relations, through which an age could be independently estimated from the stellar rotation rate. Also, we could not detect any youth indicators basing on available low SNR spectra, and GPX-1 is not presented in any X-ray database. Some methods, such as lithium abundances, require additional high SNR spectra.

We checked whether the GPX-1 system was a member of the nearby open cluster Trumpler~2 \citep{2012yCat..35430156K}, which has an age estimate of $\sim$89\,Myr obtained by \citet{2006A&A...451..901F}, $\sim$84\,Myr by \citet{2013A&A...558A..53K} and $\sim$92\,Myr by \citet{2019A&A...623A.108B} using Gaia DR2 data.  Trumpler~2 has an angular radius of 0.45\,deg and which is located at the same distance as GPX-1 (670\,pc). \citet{2012A&A...543A.156K} rated the probability of GPX-1 membership in that cluster according to its angular distance as 0\%, according to proper motion as 4.7\%, and according to 2MASS photometry as 90\% and 100\% for $J-K_{\mathrm{s}}$ and $J-H$ respectively. In addition, we tested if GPX-1 was a member of Trumpler~2 using Gaia DR2 parallaxes, proper motions and RVs. We derived cluster parameters as a local maximum in a 6-D space (Ra, Dec, distance, pmRa, pmDec, RV): Ra~=~39.32634$^{\circ}$, Dec~=~55.93867$^{\circ}$, distance~=~711\,pc, $\mathrm{pmRa~=~1.60\,mas/y}$, $\mathrm{pmDec~=~-5.27\,mas/y}$, $\mathrm{RV~=~-4.77\,km/s}$. Corresponding  parameters of GPX-1 are presented in Table~\ref{tab:coordinates}. There is a separation of 60\,pc at the present time, and the minimum separation of 27\,pc will occur in 4.5\,Myr, whereas a linear radius of Trumpler~2 for its distance and angular radius is 5.6\,pc. Thus, we believe that GPX-1 was never a member of Trumpler~2. The same conclusion about membership of the GPX-1 in Trumpler~2 was obtained in \citet{2020A&A...633A..99C}, using UPMASK procedure applied to the Gaia DR2 astrometric data \citep{2018A&A...618A..93C}.

\begin{figure*}
    \centering
    \includegraphics[width=\linewidth,trim=70 65 70 330,clip]{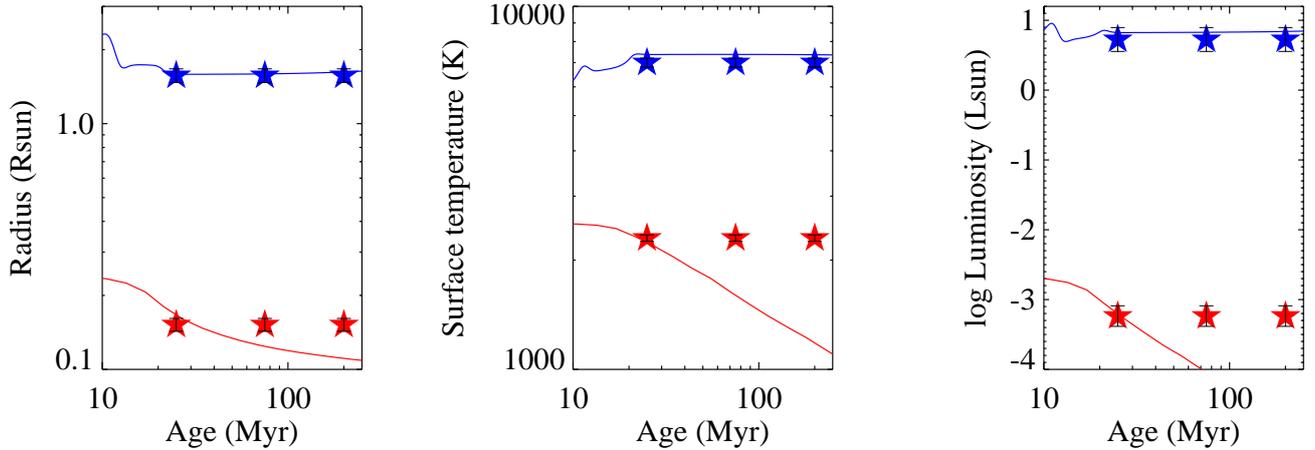}
    \caption{Evolutionary state of the GPX-1 system. In each panel, the blue curve represents the PARSEC (v1.2S) evolutionary track for the GPX-1 primary star \citep{2017ApJ...835...77M}, assuming the mass and [Fe/H] from the global solution; red curves represent the same for the GPX-1\,b brown dwarf, except using the sub-stellar evolutionary models of \citet{2020A&A...637A..38P}. Blue (red) symbols represent the observed stellar (brown dwarf) properties at three ages: 200\,Myr (close to the median estimate from our global solution); 75\,Myr; and 25\,Myr (representing the best fit to both the components simultaneously).}
    \label{fig:age_tracks}
\end{figure*}

We used the basic physical properties of GPX-1\,b and its host star together to infer the evolutionary state of the system. Figure~\ref{fig:age_tracks} shows the expected evolution of both the star and BD in radius, $T_{\rm eff}$, and luminosity, based on their masses and metallicity. The stellar track is from the PARSEC models (v1.2S; \citealt{2017ApJ...835...77M}), whereas the BD track is from the sub-stellar models of \citet{2020A&A...637A..38P}. They are compared with the measured/inferred properties at three different ages: 200\,Myr (close to the median estimate from our global solution), 75\,Myr, and 25\,Myr (chosen to match the observed properties of both the star and BD simultaneously). 

Taken at face value, all of the BD's properties are consistent with it being in an early stage of contraction at $\sim$25\,Myr, requiring no inflation. At that age, the properties of the host star are also simultaneously consistent with expectation. To be clear, the $T_{\rm eff}$ adopted here for the BD is its equilibrium temperature $T_{\rm eq}$ from the global solution, which might not be the same as the $T_{\rm eff}$ that is predicted by the models. Alternatively, an age of 75\,Myr could be correct, in which case the BD radius is somewhat inflated, but much less so than for an age of 200\,Myr.

\begin{figure}
\includegraphics[width=1.0\columnwidth]{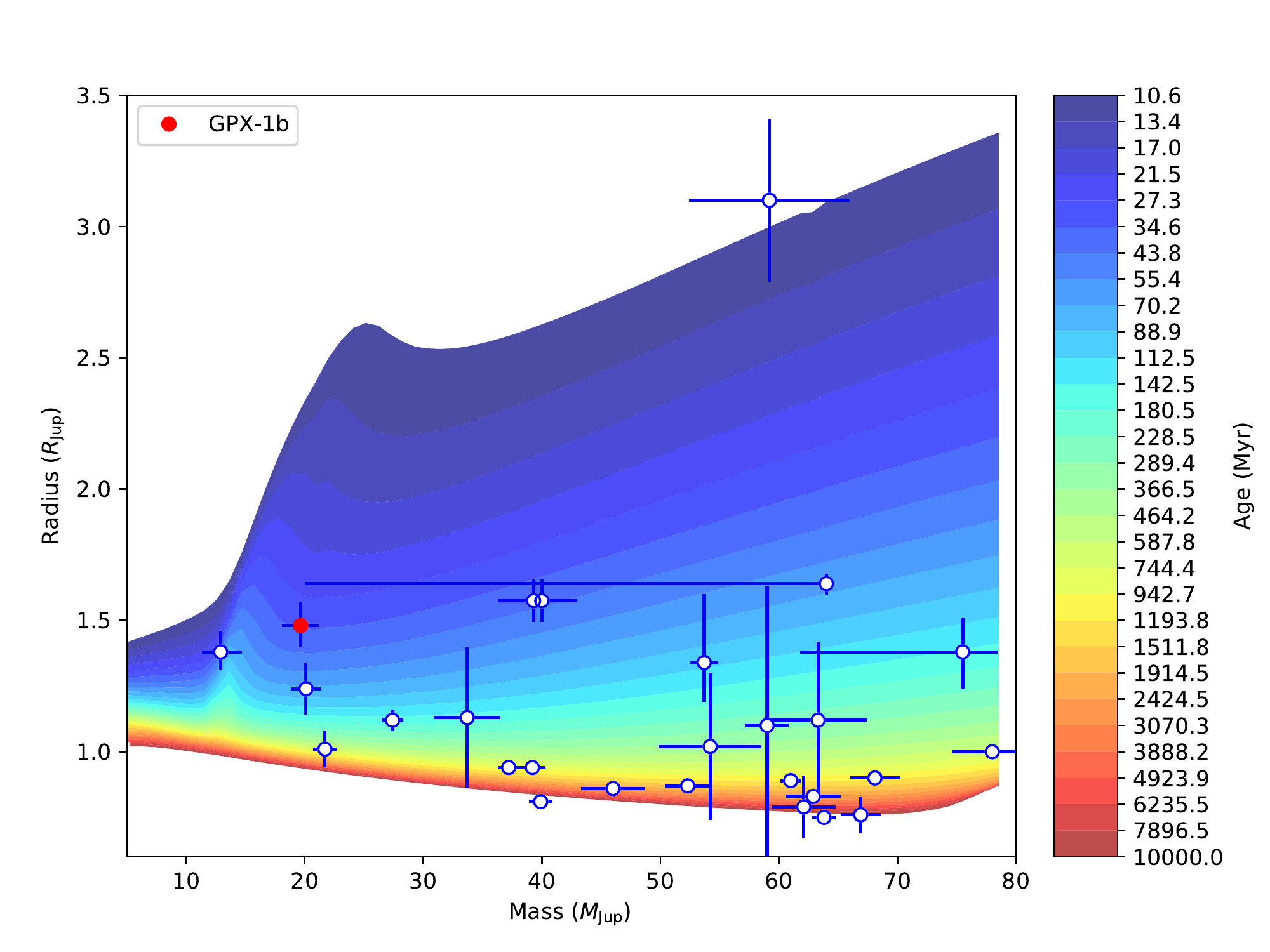}
\caption{Mass-radius relationship of a set of known transiting brown dwarfs ($\sim$13-80\,$M_{\mathrm{Jup}}$) as of May 2020 and isochrones from the evolutionary models of \citet{2020A&A...637A..38P}.} 
\label{fig:R(M)}
\end{figure}

Regarding its position on the mass-radius diagram of known transiting BDs (see Fig.~\ref{fig:R(M)}), GPX-1\,b is among the largest and youngest BDs that happen to transit their host stars. If its true age is close to a value of $0.27_{-0.15}^{+0.09}$\,Gyr obtained from our global modelling, then it is significantly inflated. Then, such an object could serve as a testbed for radius inflation theories.

We expect GPX-1\,b to have a low projected obliquity as massive exoplanets and BDs tend to be aligned \citep{2011A&A...533A.130H,2018haex.bookE...2T}, but Rossiter-McLaughlin observations should confirm this. Such a BD with measured obliquity could be important for further understanding of the origins of short-period BDs.

Assuming that the host star and the BD are black bodies, and that the thermal energy dominates, measuring the observed occultation depth can reveal the BD's effective temperature. Alternatively, we can assess expected occultation depth if $T_{\rm eff}$ of GPX-1\,b is close to its $T_{\rm Eq}$ from the global solution. In this case, occultation depth is expected to be 0.025\%, 0.05\% and 0.09\% for $J$, $H$ and $K_{\mathrm{s}}$ bands respectively. In the case of no heat redistribution, occultation depth is expected to be 0.08\%, 0.12\% and 0.17\% for $J$, $H$ and $K_{\mathrm{s}}$ bands respectively.

Given a projected rotational velocity $v\sin{\,i_*}=40\pm10$\,km/s and assuming that we observe GPX-1 edge on, the rotation period of GPX-1 appears to be synchronised with the BD orbit within $1-2\,\sigma$. The system will be in a pseudo-equilibrium state if $L > L_{\mathrm{crit}}$ , where $L$ is the total angular momentum of the system and $L_{\mathrm{crit}}$ is the critical angular momentum defined as \citep{2015A&A...574A..39D,2016A&A...589A..55D}:
$$
L_{\mathrm{crit}} = 4 \left( \frac{G^2}{27} \frac{M_\star^3 M_{\mathrm{BD}}^3}{M_\star + M_{\mathrm{BD}}} (\beta C_\star +C_{\mathrm{BD}}) \right)^{1/4},
$$

where $G$ is the gravitational constant, $M_\star$ is the stellar mass, $M_{\mathrm{BD}}$ is the mass of the BD, $C_\star$ is the moment of inertia of the star, $C_{\mathrm{BD}}$ is the moment of inertia of the BD. Parameter $\beta$ is defined as ${\Omega_\star}/{n_{\mathrm{BD}}}$, where $\Omega_\star$ is the angular velocity of the star, $n_{\mathrm{BD}} = (G (M_\star + M_{\mathrm{BD}})/a^3)^{1/2}$ is the mean motion of the BD. In the case of orbital synchronisation $\beta = 1$, meaning that the GPX-1 system is in a pseudo-equilibrium state as $L > L_{\mathrm{crit}}$.

We also assessed stability of the system and found it to be stable as $L_{\mathrm{orb}} > (4 - \beta)(C_\star + C_{\mathrm{BD}})n_{\mathrm{BD}}$, where $L_{\mathrm{orb}}$ is the orbital angular momentum defined as: 
$$
L_{\mathrm{orb}} = \left( G \frac{M_\star^2 M_{\mathrm{BD}}^2}{M_\star + M_{\mathrm{BD}}} a(1 - e^2) \right)^{1/2},
$$
where $a$ is the semi-major axis, $e$ is the orbital eccentricity of the BD.

We assessed the time needed for the system to synchronise using the work of \citet{2009MNRAS.395.2268B}. Value of tidal spin-up time $\tau_\Omega$ using the in-spiral time of the orbit $\tau_a$, shows that $\tau_\Omega$ does not exceed 250\,Myr, if the stellar tidal dissipation parameter $Q^{\prime}_\star < 2 \cdot 10^7$ (which supports our age estimate of $270_{-150}^{+90}$\,Myr from the isochrone placement algorithm). It means that during the system lifetime, it can go into a synchronized state due to tidal interaction.

\section{Conclusions}\label{sec:conclusions}

We presented here the discovery of GPX-1\,b, a transiting BD on a short circular orbit with a mass of $19.7\pm 1.6$\,$M_{\mathrm{Jup}}$ and a radius of $1.47\pm0.10$\,$R_{\mathrm{Jup}}$. The BD transits a moderately bright fast-rotating F-type star with a projected rotational velocity $v\sin{\,i_*}=40\pm10$\,km/s. Due to the small number of spectral absorption lines and their broadening, we used an isochrone placement algorithm \citep{2015A&A...575A..18B,2016A&A...585A...5B} to perform stellar characterization of GPX-1. We obtained stellar effective temperature $T_{\mathrm{eff}} = 7000\pm200$\,K, mass $1.68\pm0.10$\,$M_{\astrosun}$, radius $1.56\pm0.10$\,$R_{\astrosun}$ and approximate age $270_{-150}^{+90}$\,Myr. GPX-1 will benefit from additional spectroscopic observations with high SNR (> 100) to obtain more precise value of its projected rotational velocity and to search for possible youth indicators.

We used the basic physical properties of GPX-1\,b and its host star together to infer the evolutionary state of the system. An age of $\sim$25\,Myr matches the observed properties of both the star and BD simultaneously. In this case, no radius inflation is required. If the true age is 75\,Myr, then the BD radius is inflated, but much less than that for an age of 200\,Myr. We checked whether the GPX-1 system was a member of the nearby open cluster Trumpler~2 \citep{2012yCat..35430156K} and we believe that GPX-1 was never a member of Trumpler~2.

We find the rotation period of GPX-1 to be synchronised with the BD orbit within $1-2\,\sigma$. The system is also in a pseudo-equilibrium state and stable. The value of tidal spin-up time does not exceed 250\,Myr, which supports our age estimate from the isochrone placement algorithm ($270_{-150}^{+90}$\,Myr).

Since GPX-1 was not observed by TESS with 2-min integrations, the SPOC pipeline did not extract photometric flux nor attempted a transit search. The light curve of GPX-1 was extracted by the QLP, but no transit search was carried out as QLP reports exoplanet candidates only to TESS mag of 10.5 (and TESS mag of GPX-1 is 11.9). Transit recovery of GPX-1\,b in the TESS data is complicated by the blending by the nearby bright star HD\,15691. The same star blending issue is applicable to the WASP photometry, which prevents robust transit detections in the data. These facts provide a proof of concept of the GPX survey's scientific value and motivates us to continue its operations. The TESS mission brings exoplanet hunting into a new era with the delivery of very high-precision photometry of bright stars across the sky, beyond the ability of most ground-based telescopes. TESS will survey most of the sky over two years with thousands of expected planet discoveries \citep{2018ApJS..239....2B}. However, a number of transiting gas giants and BDs may still be discovered in crowded fields (including open clusters) when observed by ground-based surveys, and even in the TESS data, they might be difficult to detect (see Figure 3 and 4 in \citealt{2018ApJS..239....2B}). Thus, the GPX survey and similar set ups might find a niche in the TESS era and contribute by discovering new transiting gas giants and BDs in crowded fields.

\section*{Acknowledgements}
\label{Ack}
Authors would like to thank the anonymous reviewer for their time and attention. The constructive comments we received, helped us to improve the quality of the paper.

This research has made use of the Exoplanet Orbit Database, the Exoplanet Data Explorer at exoplanets.org, Extrasolar Planets Encyclopaedia at exoplanets.eu and the NASA Exoplanet Archive, which is operated by the California Institute of Technology under contract with the National Aeronautics and Space Administration under the Exoplanet Exploration Program. This research made use of Aladin \citep{2000A&AS..143...33B}. IRAF is distributed by the National Optical Astronomy Observatory, which is operated by the Association of Universities for Research in Astronomy, Inc., under cooperative agreement with the National Science Foundation. This research made use of Astropy,\footnote{http://www.astropy.org} a community-developed core Python package for Astronomy \citep{astropy:2013, astropy:2018}. 

We acknowledge the use of TESScut.MAST data from full frame time series images (FFI) collected by the TESS mission, which are publicly available from the Mikulski Archive for Space Telescopes (MAST). Funding for the TESS mission is provided by NASA’s Science Mission directorate. Resources supporting this work were provided by the NASA High-End Computing (HEC) Program through the NASA Advanced Supercomputing (NAS) Division at Ames Research Center for the production of the SPOC data products.

Paul Benni thanks Bruce Gary, the XO survey, and the KELT survey for furthering his education in exoplanet research. A.Y.B. would like to thank Catarina Fernandes and Julien de Wit for helpful discussions about the system. Organization of the EXPANSION project (E.~Sokov), follow-up campaign of the photometry observations, speckle-interferometry observations with 6-m telescope BTA were supported by the Russian Science Foundation grant 19-72-10023. The work of V.K. was supported by the Ministry of science and higher education of Russian Federation, topic no. FEUZ-0836-2020-0038. This work was partly supported by the Ministry of Science and High Education (project no. FZZE-2020-0024) and Irkutsk State University (project no. 111-14-306). This work was partially supported by the Ministry of Science and Higher Education of the Russian Federation (project no. FEUZ-2020-0030, no. 075-15-2020-780). TRAPPIST-North is a project funded by the University of Liege, in collaboration with Cadi Ayyad University of Marrakech (Morocco). E.J and M.G are F.R.S.-FNRS Senior Research Associates. The research leading to these results has received funding from the ARC grant for Concerted Research Actions financed by the Wallonia-Brussels Federation and from the Balzan Prize Foundation. TRAPPIST is funded by the Belgian Fund for Scientific Research (Fond National de la Recherche Scientifique, FNRS) under the grant FRFC 2.5.594.09.F. Erika Pak{\v s}tien{\. e} acknowledges the Europlanet 2024 RI project funded by the European Union's Horizon 2020 Research and Innovation Programme (Grant agreement No. 871149). A.B. acknowledge the support from the Program of development of M.V. Lomonosov Moscow State University (Leading Scientific School 'Physics of stars, relativistic objects and galaxies'). O.B. thanks T\"UB\.{I}TAK National Observatory for a partial support in using the T100 telescope with the project number 19AT100-1346. O.D.S.D. is supported by Portuguese national funds through Funda\c{c}\~{a}o para a Ci\^encia e Tecnologia (FCT) in the form of a work contract (DL 57/2016/CP1364/CT0004), institutional funds UIDB/04434/2020, UIDP/04434/2020 and scientific projects funds PTDC/FIS-AST/28953/2017, POCI-01-0145-FEDER-028953.

%The work of E.~D.~Kuznetsov was supported by the Ministry of science and higher education of Russian Federation under the grant 075-15-2020-780 (no. 13.1902.21.0039)

\section*{Data availability}
\label{Data availability}

Data available on request.  The data underlying this article will be shared on reasonable request to the corresponding author.
%%%%%%%%%%%%%%%%%%%%%%%%%%%%%%%%%%%%%%%%%%%%%%%%%%



\bibliographystyle{mnras}
\bibliography{bibliography}

%%%%%%%%%%%%%%%%%%%%%%%%%%%%%%%%%%%%%%%%%%%%%%%%%%

%%%%%%%%%%%%%%%%% APPENDICES %%%%%%%%%%%%%%%%%%%%%
\phantom{invisible text}\\
\textbf{Affiliations}\\
\newline
% List of institutions
$^{1}$Acton Sky Portal (Private Observatory), Acton, MA, USA\\
$^{2}$Department of Earth, Atmospheric and Planetary Sciences, Massachusetts Institute of Technology, 77 Massachusetts Avenue, Cambridge, MA 02139, USA\\
$^{3}$Instituto de Astrof\'isica de Canarias, V\'ia L\'actea s/n, 38205 La Laguna, Tenerife, Spain\\
$^{4}$Laboratory of Astrochemical Research, Ural Federal University, 620002, Mira Street, 19, Yekaterinburg, Russian Federation\\
$^{5}$Space Research Institute, Austrian Academy of Sciences, Schmiedlstrasse 6, A-8042 Graz, Austria\\
$^{6}$Space sciences, Technologies and Astrophysics Research (STAR) Institute, Universit{\'e} de Li{\`e}ge, All{\'e}e du 6 Ao{\^u}t 17, 4000 Li{\`e}ge, Belgium\\
$^{7}$Institut d'Astrophysique de Paris, UMR7095 CNRS, Universit\'e Pierre \& Marie Curie, 98bis boulevard Arago, 75014 Paris, France\\
$^{8}$Observatoire de Haute-Provence, CNRS, Universit\'e d'Aix-Marseille, 04870 Saint-Michel-l'Observatoire, France\\
$^{9}$Universit\'e Grenoble Alpes, CNRS, IPAG, 38000 Grenoble, France\\
$^{10}$Instituto de Astrof{\'\i}sica e Ci\^encias do Espa\c{c}o, Universidade do Porto, CAUP, Rua das Estrelas, 4150-762 Porto, Portugal\\
$^{11}$INAF -- Osservatorio Astrofisico di Arcetri, Largo E. Fermi 5, 50125, Firenze, Italy\\
$^{12}$Department of Physics, Lehigh University, 16 Memorial Drive East, Bethlehem, PA 18015, USA\\
$^{13}$Department of Physics and Astronomy, Vanderbilt University, 6301 Stevenson Center Ln., Nashville, TN 37235, USA\\
$^{14}$Department of Astronomy, The University of Texas at Austin, Austin, TX 78712, USA\\
$^{15}$NASA Sagan Fellow\\
$^{16}$Sternberg Astronomical Institute, M.V. Lomonosov Moscow State University, 13, Universitetskij pr., 119234 Moscow, Russia\\
$^{17}$Faculty of Physics, M.V. Lomonosov Moscow State University, Leninskie Gory, 1, 119991, Moscow, Russia\\
$^{18}$Astrobiology Research Unit, Universit\'e de Li\`ege, All\'ee du 6 Ao\^ut 19C, 4000 Li\`ege, Belgium\\
$^{19}$Oukaimeden Observatory, High Energy Physics and Astrophysics Laboratory, Cadi Ayyad University, Marrakech, Morocco\\
$^{20}$National Youth Space Center, Goheung, Jeollanam-do, 59567, S.~Korea\\
$^{21}$Federal State Budget Scientific Institution Crimean Astrophysical Observatory of RAS, Nauchny, 298409, Crimea, Russia\\
$^{22}$Special Astrophysical Observatory, Russian Academy of Sciences, Nizhnij Arkhyz, Russia, 369167\\
$^{23}$Saint Petersburg State University, Faculty of Mathematics \& Mechanics, Universitetskij pr. 28, Petrodvorets, St. Petersburg 198504, Russia\\
$^{24}$Ural Federal University, 620002, Mira Street, 19, Yekaterinburg, Russian Federation\\
$^{25}$Centre for Exoplanets and Habitability, University of Warwick, Gibbet Hill Road, Coventry CV4 7AL, United Kingdom\\
$^{26}$Department of Physics, University of Warwick, Gibbet Hill Road, Coventry CV4 7AL, United Kingdom\\
$^{27}$Department of Physics, and Kavli Institute for Astrophysics and Space Research, Massachusetts Institute of Technology, Cambridge, MA 02139, USA\\
$^{28}$Center for Astrophysics \textbar \ Harvard \& Smithsonian, 60 Garden St., Cambridge, MA 02138, USA\\
$^{29}$Department of Aeronautical and Astronautical Engineering, Massachusetts Institute of Technology, Cambridge, MA, 02139\\
$^{30}$NASA Ames Research Center, Moffett Field, CA 94035, USA\\
$^{31}$Central Astronomical Observatory at Pulkovo of Russian Academy of Sciences, Pulkovskoje shosse d. 65, St. Petersburg, Russia, 196140\\
$^{32}$Astronomical Observatory - DSFTA, University of Siena, Via Roma 56, 53100 Siena, Italy\\
$^{33}$Wild Boar Remote Observatory, San Casciano in Val di Pesa (FI), Italy\\
$^{34}$Ankara University, Faculty of Science, Department of Astronomy and Space Science, TR-06100 Tandogan, Ankara, Turkey\\
$^{35}$Irkutsk State University, K. Marx str., 1, Irkutsk, 664003, Russia\\
$^{36}$Institute of Theoretical Physics and Astronomy, Vilnius University, Saul\.etekio al. 3, Vilnius, LT-10257, Lithuania\\
$^{37}$Taurus Hill Observatory, H\"{a}rk\"{a}m\"{a}entie 88, 79480 Varkaus, Finland\\
$^{38}$Physics and Engineering Physics Department, University of Saskatchewan, Saskatoon, SK, Canada, S7N 5E2\\
$^{39}$Baronnies Proven\c{c}ales Observatory, Hautes Alpes - Parc Naturel R\' egional des Baronnies Proven\c{c}ales, 05150 Moydans, France\\
$^{40}$GJP private observatory, Elgin, OR, USA\\
$^{41}$Rasteau Observatory, 84110 Rasteau, France\\
$^{42}$Anunaki Observatory, Manzanares El Real, Spain\\
$^{43}$Grand-Pra private observatory, Switzerland\\
$^{44}$Observatory Ca l’Ou, Sant Mart{\'i} Sesgueioles, Spain\\
$^{45}$Observatori Montcabrer, Spain\\
$^{46}$Department of Physics and Astronomy, Brigham Young University, Provo, UT 84602 USA\\
$^{47}$Center for Exoplanets and Habitable Worlds, The Pennsylvania State University, 525 Davey Lab, University Park, PA 16802, USA\\
$^{48}$Department of Astronomy \& Astrophysics, The Pennsylvania State University, 525 Davey Lab, University Park, PA 16802, USA\\
$^{49}$Eberly Research Fellow\\
$^{50}$Swarthmore College Dept. of Physics \& Astronomy, 500 College Ave., Swarthmore PA 19081 USA\\
$^{51}$NASA Goddard Space Flight Center, 8800 Greenbelt Rd, Greenbelt, MD 20771\\
$^{52}$Space Telescope Science Institute, 3700 San Martin Drive, Baltimore, MD, 21218, USA\\

%%%%%%%%%%%%%%%%%%%%%%%%%%%%%%%%%%%%%%%%%%%%%%%%%%

% Don't change these lines
\bsp	% typesetting comment
\label{lastpage}
\end{document}